\documentclass[reprint,aps,pra, 
superscriptaddress,
 amsmath,amssymb,
]{revtex4-2}

\usepackage{graphicx}
\usepackage{dcolumn}
\usepackage{bm}
\usepackage{float}
\usepackage{subcaption}
\usepackage{hyperref}
\usepackage{ragged2e}
\usepackage{caption}

\usepackage{comment}
\begin{document}

\preprint{APS/123-QED}

\title{Single-photon addition to multimode quantum fields at telecom wavelengths using lithium niobate and its doped variants}

\author{Naveen V}
 \affiliation{School of Quantum Technology, Defence Institute of Advanced Technology, Pune 411024, India.}

\author{Nicolas Treps}
\affiliation{Laboratoire Kastler Brossel, Sorbonne Universit\'{e}, CNRS, ENS-PSL Research University, Coll\`{e}ge de France, 4 place Jussieu, F-75252 Paris, France}

\author{Raghavan G}%
\affiliation{School of Quantum Technology, Defence Institute of Advanced Technology, Pune 411024, India.}%

\author{Valentina Parigi}
\affiliation{Laboratoire Kastler Brossel, Sorbonne Universit\'{e}, CNRS, ENS-PSL Research University, Coll\`{e}ge de France, 4 place Jussieu, F-75252 Paris, France}

\author{Srinivasan K}%
 \email{Corresponding author: srinik@diat.ac.in}
\affiliation{School of Quantum Technology, Defence Institute of Advanced Technology, Pune 411024, India.}%

\date{\today}

\begin{abstract}

Multimode spectro-temporal non-Gaussian quantum states of light play a crucial role in continuous-variable quantum information processing. In particular, the non-Gaussian quantum states generated through a single-photon addition (SPA) to multimode quantum fields at telecommunication wavelengths constitute a useful resource for quantum technologies. We present a detailed theoretical analysis and numerical simulations of single-photon addition (SPA) and mode-selective single-photon addition (MSSPA) to a multimode quantum field at telecommunication wavelengths. The analysis is performed for congruently grown lithium niobate (LN), $5\%$ MgO-doped lithium niobate (MgO:LN) and various concentrations of ZnO-doped lithium niobate (ZnO:LN) bulk crystals. We investigate MSSPA using type-II parametric down-conversion (PDC) and SPA using the type-I PDC processes. We further present spectro-temporal mode-selective SPA at the telecom wavelength using $5\%$ MgO:LN for temperatures ranging from $0^\circ C$ to $15^\circ C$. In addition, we perform simulations to investigate SPA based on the noncollinear type-I PDC configuration. These theoretical results identify optimal experimental conditions and phase-matching parameters to achieve SPA and mode-selective SPA for LN-based crystals, providing a potential platform for non-Gaussian quantum state engineering at telecommunication wavelengths, mainly at $1550$ nm.
\end{abstract}
 
\maketitle

\section{\label{sec:level1}Introduction}

Non-Gaussian quantum states of light play a major role in continuous-variable quantum information processing because the resources provided by these states cannot be achieved using Gaussian quantum states and Gaussian quantum operations alone ~\cite{PRXQuantum2021,Ra2020NonGaussian}. The multimode spectro-temporal non-Gaussian quantum states of light, such as photon-added or photon-subtracted multimode quantum states, are important resources for continuous variable quantum communication, as they offer a high-dimensional Hilbert space for encoding information~\cite{Walschaers2017Statistical,Ra2019Non-Gaussian}. The negative values of the Wigner distribution of these non-Gaussian states are proven to be a valid resource for quantum computation~\cite{Walschaers2017Entanglement,Walschaers2021Non-Gaussian,agarwal1991nonclassical,namekata2010non,ourjoumtsev2006generating,wenger2004non,kim2005nonclassicality}. The experimental generation of such states is vital for the implementation of scalable and controllable quantum computing.

The non-Gaussian spectro-temporal quantum states are less influenced by the environment compared to other degrees of freedom of photons, such as the polarization and orbital angular momentum states of light~\cite{Brecht2015Photon}. This property places them as a very suitable candidate for continuous variable quantum communication and computation. Among all non-Gaussian quantum states generated from the spectro-temporal modes, the single-photon added or subtracted multimode quantum fields have been demonstrated and being a prominent resource in recent years~\cite{Endo2023Non-Gaussian,Chen2024Generation}. Mainly, the focus is on the generation of SPA to multimode quantum fields at telecommunication wavelengths, as the optical fibers exhibit minimal loss at these wavelength region~\cite{Dibos2017Atomic,Xu2022A} and the commonly used method to experimentally achieve this is using parametric down-conversion (PDC)~\cite{zavatta2004quantum,zavatta2005single,zavatta2007experimental,parigi2007probing,parigi2008manipulating,fadrny2024experimental,Chen2024Generation}.
The present work focuses on tailoring the spectro-temporal modes of light to generate non-Gaussian quantum states such as single photon-added multimode quantum states at telecom wavelength. We present our theoretical and simulation results on the generation of single-photon-added multimode quantum fields using a lithium niobate and its doped variants, namely, MgO and ZnO.

A theoretical framework for the generation of mode-selective addition of photons to a multimode quantum states using the type-II collinear PDC process has been proposed earlier~\cite{roeland2022mode}. This theoretical study demonstrated that the necessary condition for the spectro-temporal mode-selective addition is to match the group velocities of the pump and the signal photons. Once the group velocities are matched, the careful selection of the crystal length and the pump bandwidth enables that the generated spectro-temporal mode of the signal matches the spectral mode of the pump. This makes it possible to tailor the spectro-temporal modes of light for realizing the mode-selective non-Gaussian quantum operations such as single photon addition. The theoretical results were demonstrated for a variety of uniaxial and biaxial nonlinear crystals. It showed that the mode-selective addition of photons is not possible with a congruently grown lithium niobate crystal at room temperature, as the phasematching and group velocity matching conditions cannot be simultaneously satisfied. In the present work, we extend this study to lithium niobate (LN), MgO-doped anZnO-dopeded lithium niobate crystals at various crystal temperatures. We identify that mode-selective addition is possible with the $5\%$ MgO doped lithium niobate crystal within the temperature range of $0^\circ C$ to $15^\circ C$. The group velocity matching and the refractive index matching curves intersect at the pump wavelength of $778$ nm at $15^\circ$C. This enables the generation of mode-selective addition of photons using $5\%$ MgO:LN crystals at telecom wavelength of $1556 ~ nm$. The possibility of realizing Type-II PDC in ZnO:LN is investigated for various ZnO concentrations and crystal temperatures. We find that a Type-II PDC is not achievable in ZnO:LN. We present an analytical expression and simulations for the spectro-temporal mode-selective SPA at telecom wavelength.

We extend our theoretical study to the noncollinear type-I PDC process in LN, MgO:LN and ZnO:LN crystals for various crystal temperatures. Earlier studies have shown that, in the case of non-collinear Type-I PDC process, only SPA is possible, whereas mode-selective SPA is not achievable in this configuration. As mentioned earlier, the necessary condition for the SPA is the group velocity matching between the pump and the signal photons. For a given crystal, group velocity and the refractive index matching curves intersect exactly at one specific pump wavelength. This restricts the generation of SPA at arbitrary wavelengths. In the present work, we employ the approach proposed by U'Ren et al.~\cite{u2003photon} to generate the spectrally pure single photons without requiring the group velocity matching condition. With this theoretical approach single photon added multimode quantum states can be generated at arbitrary wavelengths, wherever the refractive index matching is achievable. This enables the realization of SPA using the noncollinear type-I PDC process at telecom wavelength. However, when the group velocities of the pump and signal are matched, the conditions required by this approach lead to an unrealistic experimental scenario. To overcome this limitation, we provide an analytical expression that remains valid under the GVM conditions. Therefore, when both the GVM and the analytical conditions are achieved together, SPA with very high spectral purity can be realized. This procedure guarantees that the effective number of spectral modes present in the signal field is exactly one.  We present simulation results for SPA using LN, MgO:LN and ZnO:LN crystals. The primary focus of this study is to generate SPA to a multimode quantum states at telecom wavelengths. We demonstrate that the condition derived in this work enables the realization of SPA with high spectral purity even though the GVM is achieved. 

The present work is organized as follows. Section II presents the general theory for the single-photon addition to multimode quantum fields. Section III presents a theory for the mode-selective single photon addition using the collinear type - II PDC process, and the corresponding simulation results are given in section IV. The theory and the numerical simulations for the single-photon addition using the noncollinear type - I PDC process are presented in section V and VI. The discrepancies in the reported refractive indices of ZnO-doped lithium niobate are discussed in Section VII. Finally, a summary of the work and the conclusions are presented in Section VIII.

\section{Theory of single photon addition to a multimode quantum field using parametric downconversion (PDC) process}
\begin{figure}[b]
\includegraphics[width=0.45\textwidth]{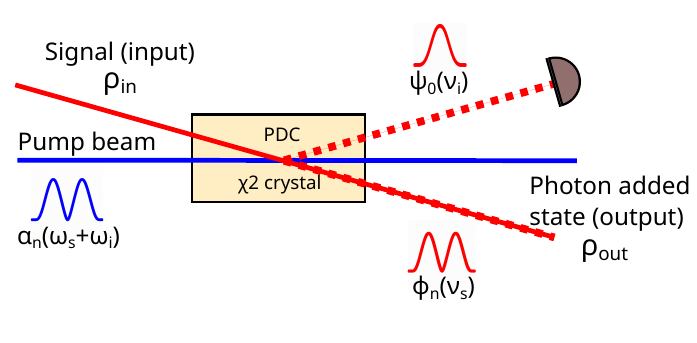}
\caption{\label{fig:SPA Schematic} Schematic illustration of mode-selective single-photon addition to a multimode quantum field $\hat{\rho}$. The multimode quantum field $\hat{\rho}$ and a pump field in the $\mathrm{HG}_n$ mode interact with a nonlinear crystal, generating a single-photon-added multimode quantum field. The photon-addition process is heralded by the detection of a photon in the idler mode.
}
\end{figure}

In the present work, we derive an analytical expression for JSA for the single-photon addition (SPA) to a multimode quantum field using the PDC process. We present simulation results for LN and its doped variants. The realization of SPA requires the generation of spectrally uncorrelated signal-idler pairs, which can be achieved by tailoring the joint spectral amplitude of the PDC process. This involves factorizing the joint spectral amplitude (JSA) into individual signal and idler frequency components. By appropriately tailoring the JSA, the spectro-temporal mode generated in the signal field can be matched to that of a pump beam, thereby enabling mode-selective photon addition to a multimode quantum field.

Consider a spectro-temporal multimode quantum field with a quantum state $\hat{\rho}$. SPA is achieved with the help of a nonlinear optical PDC process using nonlinear crystals. The multimode quantum field interacts with the pump field in the nonlinear medium, resulting in the generation of signal and idler photons. The signal photons are added to the multimode quantum field $\hat{\rho}$ to generate the SPA to a multimode field. The detection of idler photons heralds the generation of the SPA, as illustrated in Fig.~\ref{fig:SPA Schematic}. 

The initial state of the signal-idler spectro-temporal mode is given by $|0\rangle_s \otimes |0\rangle_i$. The creation and annihilation operators associated with signal and idler spectro-temporal modes are $\{\hat{a}^{\dagger}_s , \hat{a}_s\}$ and $\{\hat{a}^{\dagger}_i , \hat{a}_i\}$ respectively. The Hamiltonian associated with PDC process is given by~\cite{walborn2010spatial},

\begin{equation}
    \hat{H} = \int \int d \omega_s d\omega_i J(\omega_s,\omega_i) \hat{a}^{\dagger}_s \hat{a}^{\dagger}_i + h. c
\end{equation}

where $J(\omega_s,\omega_i)$ is the joint spectral amplitude (JSA) of the PDC process, which describes the spectral correlation between the down-converted signal-idler photon pair. JSA is the product of the pump envelope function (PEF) $\alpha(\omega_s, \omega_i)$ and the phase matching function (PMF) $\phi(\omega_s, \omega_i)$~\cite{zielnicki2018joint,u2006generation}. 
\begin{equation}
    J(\omega_s, \omega_i) = \alpha(\omega_s, \omega_i) \times \phi(\omega_s, \omega_i)
\end{equation} 

 The PEF can be expressed as, 
\begin{equation}
    \alpha(\omega_s, \omega_i) = e^{- \frac{\left( \nu_{i} + \nu_{s} \right)^{2}}{2\sigma_p^{2}}}.
\end{equation}
 For the noncollinear process, the PMF itself is a product of the transverse and longitudinal phase-matching functions~\cite{u2003photon}.

\begin{equation}
    \phi(\omega_s, \omega_i) = \phi_z(\omega_s, \omega_i) \phi_{\perp}(\omega_s, \omega_i)
\end{equation}

The longitudinal and the transverse phase-matching functions are related to the momentum mismatch in the longitudinal $\Delta K_{z}$ and transverse directions $\Delta K_{\bot}$, it is given by,
\begin{equation}
    \phi_z(\omega_s, \omega_i) = e^{- \frac{\gamma\Delta K_{z}^{2}L^{2}}{4}}
\end{equation}

and 
\begin{equation}
    \phi_{\perp}(\omega_s, \omega_i) = e^{- \frac{\Delta K_{\bot}^{2}w_{o}^{2}}{4}}
\end{equation}
Therefore, the JSA can be expressed as,

\begin{equation}\label{eq:JSA}
JSA = e^{- \frac{\left( \nu_{i} + \nu_{s} \right)^{2}}{2\sigma_p^{2}}} e^{- \frac{\gamma\Delta K_{z}^{2}L^{2}}{4}} e^{- \frac{\Delta K_{\bot}^{2} w_{o}^{2}}{4}}
\end{equation}
where, $\nu_{s} = \omega_{s}- \omega_{so}$ and $\nu_{i} = \omega_{i} - \omega_{io}$ are the frequency offsets from the central frequencies $\omega_{so}$ and $\omega_{io}$ and $\gamma \approx 0.193$. Here $\sigma_p$, $L$ and $w_0$ are the pump bandwidth, crystal length and the pump beam waist, respectively. The first-order Taylor expansions of the transverse and longitudinal components of the momentum mismatch are given by
\begin{equation}
\Delta K_{z} = \Delta K_{z}^{o}
+ \left( K_{p}^{'} - K_{s}^{'}\cos\theta_{s} \right)\nu_{s}
+ \left( K_{p}^{'} - K_{i}^{'}\cos\theta_{i} \right)\nu_{i}    
\end{equation}
and 
\begin{equation}
\Delta K_{\bot} = \Delta K_{\bot}^{o}
+ \left( K_{s}^{'}\sin\theta_{s} \right)\nu_{s}
- \left( K_{i}^{'}\sin\theta_{i} \right)\nu_{i}    
\end{equation}

By the careful selection of the pump bandwidth ($\sigma_p$), crystal length($L$) and pump beam waist ($w_o$) along with the phase matching conditions, the spectral properties and the correlation between the photon pairs can be controlled. The Schmidt decomposition of the JSA gives us

\begin{equation}\label{schmidt decomp}
    JSA = \sqrt{\lambda_n}\psi_n(\nu_i)\phi_n(\nu_s)
\end{equation}
Where $\lambda_n$ is Schmidt coefficient and $\psi_n(\nu_i)$ and $\phi_n(\nu_s)$ are signal and idler spectrum of the corresponding Schmidt mode $n$.
The spectral correlation between the signal and idler photons is measured by a quantity called Schmidt number, and it is given by,
\begin{equation}
    K = \frac{1}{\Sigma\lambda_i^2}
\end{equation}

When a generated photon pair is spectrally uncorrelated, then there will be only one Schmidt coefficient that survives. Therefore, the Schmidt decomposition of its joint spectral amplitude can be written as a product of signal and idler frequency modes. This results in the generation of a spectrally pure single photon pair with K = 1.

Now going back to the joint spectral amplitude given in Eq.~\eqref{eq:JSA} and rearranging it, ends up with a product of three functions. That is, the signal component, the idler component and their correlation.

\begin{align}\label{eq: jsa expand}
JSA= e^{-a\nu_s^2}e^{-b\nu_i^2}e^{-c\nu_s\nu_i}
\end{align}

Where, 
\begin{align*}
a =&  \frac{1}{2\sigma_p^{2}} + \frac{\gamma\left( K_{p}^{'} - K_{s}^{'}\cos\theta_{s} \right)^{2}L^{2}}{4} + \frac{\left( K_{s}^{'}\sin\theta_{s} \right)^{2}w_{o}^{2}}{4}\\
b =&\frac{1}{2\sigma_p^{2}} + \frac{\gamma\left( K_{p}^{'} - K_{i}^{'}\cos\theta_{i} \right)^{2}L^{2}}{4} + \frac{\left( K_{i}^{'}\sin\theta_{i} \right)^{2}w_{o}^{2}}{4}\\
c =&\frac{1}{\sigma_p^{2}} + \frac{\gamma\left( K_{p}^{'} - K_{s}^{'}\cos\theta_{s} \right)\left( K_{p}^{'} - K_{i}^{'}\cos\theta_{i} \right)L^{2}}{2}\ \notag\\
&- \ \frac{\left( K_{s}^{'}\sin\theta_{s} \right)\left( K_{i}^{'}\sin\theta_{i} \right)w_{o}^{2}}{2}
\end{align*}
When $c = 0$, the component that establishes a correlation between signal and idler frequencies is removed~\cite{u2006generation,grice2001eliminating}. Hence, the JSA will only be a product of signal and idler components, leading to spectrally uncorrelated photon pairs. So the condition for generating a spectrally uncorrelated photon pair directly from the non-linear crystal is $c = 0$ (i.e.)

\begin{align}\label{eq: uncorrelation condition}
\frac{1}{\sigma_p^{2}} + \frac{\gamma\left( K_{p}^{'} - K_{s}^{'}\cos\theta_{s} \right)\left( K_{p}^{'} - K_{i}^{'}\cos\theta_{i} \right) L^{2}}{2}\notag
\\ -\frac{ \left( K_{s}^{'}\sin\theta_{s} \right)\left( K_{i}^{'}\sin\theta_{i} \right) w_{o}^{2}}{2} = 0 
\end{align}

The Eq.~\eqref{eq: uncorrelation condition} provides a condition for generating pure spectro-temporal modes for the signal and idler photons so that the Schmidt number K=1 for the SPA process. This is the general condition for type-0, type-I and type-II PDC processes.

\section{Theory of mode-selective SPA using Type-II collinear PDC process}
 Roeland et al.~\cite{roeland2022mode} showed that the necessary condition for achieving the mode-selective single-photon addition is group-velocity matching between the pump and the signal photons. Once this condition is satisfied, choosing a longer crystal or a broader pump bandwidth enables the generation of mode-selective SPA. In the present section, we provide an analytical expression for the mode-selective addition process. We show that, once the above-mentioned conditions are satisfied, the spectro-temporal mode of the signal matches the desired HG mode of the pump. This offers a possibility of adding a desired spectro-temporal mode to a multimode quantum field $\hat{\rho}$ in a mode-selective way.

In collinear Type-II parametric down-conversion process, a pump photon of frequency $\omega_p$ and the multimode quantum field $\hat{\rho}$ with a central frequency $\omega_s$ interact with a nonlinear crystal to generate two orthogonally polarized lower-energy signal and idler photons with central frequencies $\omega_s$ and $\omega_i$. For the collinear PDC process $\theta_s=-\theta_i=0$. Using Eq.~\eqref{eq: uncorrelation condition}, the condition for pure state generation becomes

\begin{equation}\label{uncorrelation type2}
    \frac{1}{\sigma_p^{2}} + \frac{\gamma \left( K_{p}^{'} - K_{s}^{'} \right)\left( K_{p}^{'} - K_{i}^{'} \right)L^{2}}{2} = 0
\end{equation}

Rearranging the terms in the equation gives us the following:
\begin{equation}
    \frac{2}{\sigma_p L\left( K_{p}^{'} - K_{i}^{'} \right)} +  \sigma_p L\gamma\left( K_{p}^{'} - K_{s}^{'} \right) = 0
\end{equation}

For a sufficiently long crystal or a broader pump beam, the first term would be infinitesimally small and it can be ignored. 
\begin{equation}\label{eq:long crystal broad pump}
    \delta= \frac{2}{\sigma_p L\left( K_{p}^{'} - K_{i}^{'} \right)}
\end{equation}

This results in equation $K'_p-K'_s=0$, which can also be expressed in terms of group velocities as,
\begin{equation}\label{eq:gvm type2}
    v_g^p=v_g^s
\end{equation}

As mentioned earlier, the GVM between the pump and signal photons is a necessary condition for the mode-selective SPA. The PDC process requires the phase-matching condition, which is determined by the refractive indices of the interacting fields, whereas the mode-selective SPA additionally requires the GVM condition to be satisfied. Therefore, both GVM and the index-matching conditions need to be satisfied for the SPA. However, for a given crystal, this cannot be achieved for an arbitrary wavelength of the pump beam, as GVM and the index matching curves intersect at exactly one given wavelength of the pump beam.  

 Let us assume that the pump envelope function is chosen as the $n$-th order Hermite-Gaussian mode $HG_n$. Now, the joint spectral amplitude of the down-converted photon pair is given by~\cite{roeland2023quantum},

\begin{equation}
    JSA =H_{n}\left( \frac{\left( \nu_{s} + \nu_{i} \right)}{\sigma_p} \right) e^{- \nu_{s}^{2}\left( \frac{1}{2\sigma_p^{2}} \right)}e^{- \nu_{i}^{2}\left( \frac{1}{2\sigma_p^{2}} + \frac{\gamma\left( K_{p}^{'} - K_{i}^{'} \right)^{2}L^{2}}{4} \right)}
\end{equation}

Where $HG_n$ is the $n$-th order Hermite-Gaussian function. Substituting (8) in (10) results in,

\begin{equation}\label{eq:jsa type 2 preexpansion}
 JSA = H_{n}\left( \frac{\left( \nu_{s} + \nu_{i} \right)}{\sigma_p} \right)e^{- \nu_{s}^{2}\left( \frac{1}{2\sigma_p^{2}} \right)}e^{- \frac{\nu_{i}^{2}}{2\sigma_p^{2}}\left( 1 + \frac{2\gamma}{\delta^{2}} \right)}   
\end{equation}

As given in Eq.~\eqref{eq:long crystal broad pump}, for a sufficiently longer crystal length or for a broader pump, the term $\delta$ is negligibly small. Therefore, from the third term of the Eq.~\eqref{eq:jsa type 2 preexpansion}, it is clear that the spectral distribution along the idler is a Gaussian function with a very narrow bandwidth compared to the signal spectrum. This restricts the frequency offset from the central frequency $\nu_i=\omega_i - \omega_{io}$, because after a certain value of the frequency offset, the JSA becomes zero.    

The narrow bandwidth of the idler frequency distribution makes the Gaussian envelope highly localized around $\nu_i=0$, in contrast to the slowly varying Hermite polynomial $H_n$. As $\nu_i$ moves away from the origin, the magnitude of the Gaussian envelope rapidly decreases toward zero due to its narrow bandwidth. Consequently, this makes the $\nu_i$ components that contribute most significantly to the JSA are concentrated near the centre of the Gaussian, $\nu_i=0$. In comparison with the $\nu_i$-dependent Gaussian envelope, the Hermite polynomial varies slowly with $\nu_i$. Since the dominant contributions to the JSA arise from values of $\nu_i$ close to the origin, we assume that $\nu_i/\sigma_p$ is a small quantity, denoted by $\epsilon$, such that $\nu_i/\sigma_p=\epsilon\ll1$. Under this approximation, the JSA becomes

\begin{align}
 = {\left( H_{n}\left( \frac{\nu_{s}}{\sigma_p} \right) + \epsilon H_{n}^{'}\left( \frac{\nu_{s}}{\sigma_p} \right) + \frac{\epsilon^{2}}{2!}H_{n}^{''}\left( \frac{\nu_{s}}{\sigma_p} \right) + \ldots \right)}\notag\\
 e^{- \frac{\nu_{s}^{2}}{2\sigma_p^{2}}}e^{- \frac{\nu_{i}^{2}}{2\sigma_p^{2}}\left( 1 + \frac{2\gamma}{\delta^{2}} \right)}
\end{align}


As mentioned earlier,

(i) For a longer crystal or a broader pump \((\delta \approx 0)\).

(ii) the GVM condition \(K_{p}^{'} = K_{s}^{'}\) is satisfied to
generate a spectrally uncorrelated JSA. But it also imposes a restriction on JSA,
as the HG polynomial varies slowly relative to the Gaussian
with respect to \(\nu_{i}\), hence ~\cite{roeland2022mode},

\begin{equation}\label{eq:jsa type2}
    JSA = {H_{n}\left( \frac{\nu_{s}}{\sigma_p} \right)}e^{- \frac{\nu_{s}^{2}}{2\sigma_p^{2}}}e^{- \frac{\nu_{i}^{2}}{2\sigma_p^{2}}\left( 1 + \frac{2\gamma}{\delta^{2}} \right)}
\end{equation}
As given in the previous expression for JSA (Eq.~\eqref{eq:jsa type2}), it is clear that the idler spectrum is always a Gaussian. The spectro-temporal mode of the signal is the same as the pump due to the conditions given above. Already, there exists a multimode quantum field in the signal arm. Whenever a photon is detected in the idler arm, a photon with the desired HG mode is going to be added to the multimode quantum field. In the next section we provide a detailed simulation results for various doped LN crystals.

\section{Simulation results for mode-selective single-photon addition using Type-II PDC}
In this section, we present the simulation results for the mode-selective SPA to a spectro-temporal multimode signal $\hat{\rho}$. The phase-matching condition for the Type-II collinear and degenerate PDC process is given by, 
\begin{equation}\label{eq:pm type2}
    2 n_e (\lambda_p, \theta_{PM}) = n_o (\lambda_s) + n_e (\lambda_i, \theta_{PM})
\end{equation}
As given earlier, the necessary condition for the mode-selective SPA is the group velocity matching between pump and signal photons. Roeland et al.,~\cite{roeland2022mode} showed that there is a restriction on choosing the signal polarization. As the e-polarized signal does not obey the group velocity matching condition, it is always necessary to choose the signal photon as o-polarized. The group velocity matching condition is given by,
\begin{equation}
    v_g(\lambda_p, \theta_{PM}) = v_g (\lambda_s)
\end{equation}
where the group velocities are the inverse of the first derivative of wave vector with respect to frequency, $\frac{1}{v_g} = \frac{\partial k}{\partial \omega}$. The mode-selective addition is possible if both the phase-matching and group velocity matching conditions are achieved together. However, achieving both conditions is not possible for every wavelength. Because the phase-matching and the group velocity angles are not the same for a given wavelength. The wavelength for which both conditions are possible is identified by plotting the phase-matching and group velocity matching curves for different wavelengths. The intersection point of these two curves says that both phase-matching and group velocity matching angles are the same for that specific wavelength. For the congruently grown lithium niobate crystal, this condition was verified for several wavelengths. For LN, the coefficients of the ordinary polarization were taken from ~\cite{edwards1984temperature} while the coefficients of the extraordinary polarisation were taken from ~\cite{jundt1997temperature}. For the degenerate and collinear Type-II PDC process, the phase-matching condition is possible for the wavelength range between $869$ nm to $1193$ nm. For the same configuration the group velocity matching is possible between $808$ nm to $1347$ nm. The phase-matching and the group-velocity matching curves are plotted between the wavelength range $869$ nm to $1193$ nm. For the congruently grown LN, the possibility of performing mode-selective addition has been checked for various crystal temperatures ranging from $10^\circ C$ to $100^\circ C$. But the phase matching and group velocity matching curves never intersect each other in the $\lambda-\theta_{PM}$ plane in the specified wavelength range. Therefore, as discussed in Roeland et al.~\cite{roeland2022mode}, the mode-selective addition is not possible for the LN. 

In the present work, we extended that work for MgO:LN and ZnO:LN crystals at different crystal temperatures. The Sellmeier coefficients for 5\% MgO-doped LN and ZnO-doped LN were taken from~\cite{gayer2008temperature} and~\cite{schlarb1995refractive}. For the ZnO:LN, the type-II PDC is not possible for any wavelength at any temperature. Therefore, this excludes the possibility of using LN and ZnO:LN for the mode-selective addition using the Type-II PDC process. 
\begin{figure}[b]
 \vspace{3mm}
 \begin{center}
 \includegraphics[width=0.4\textwidth]{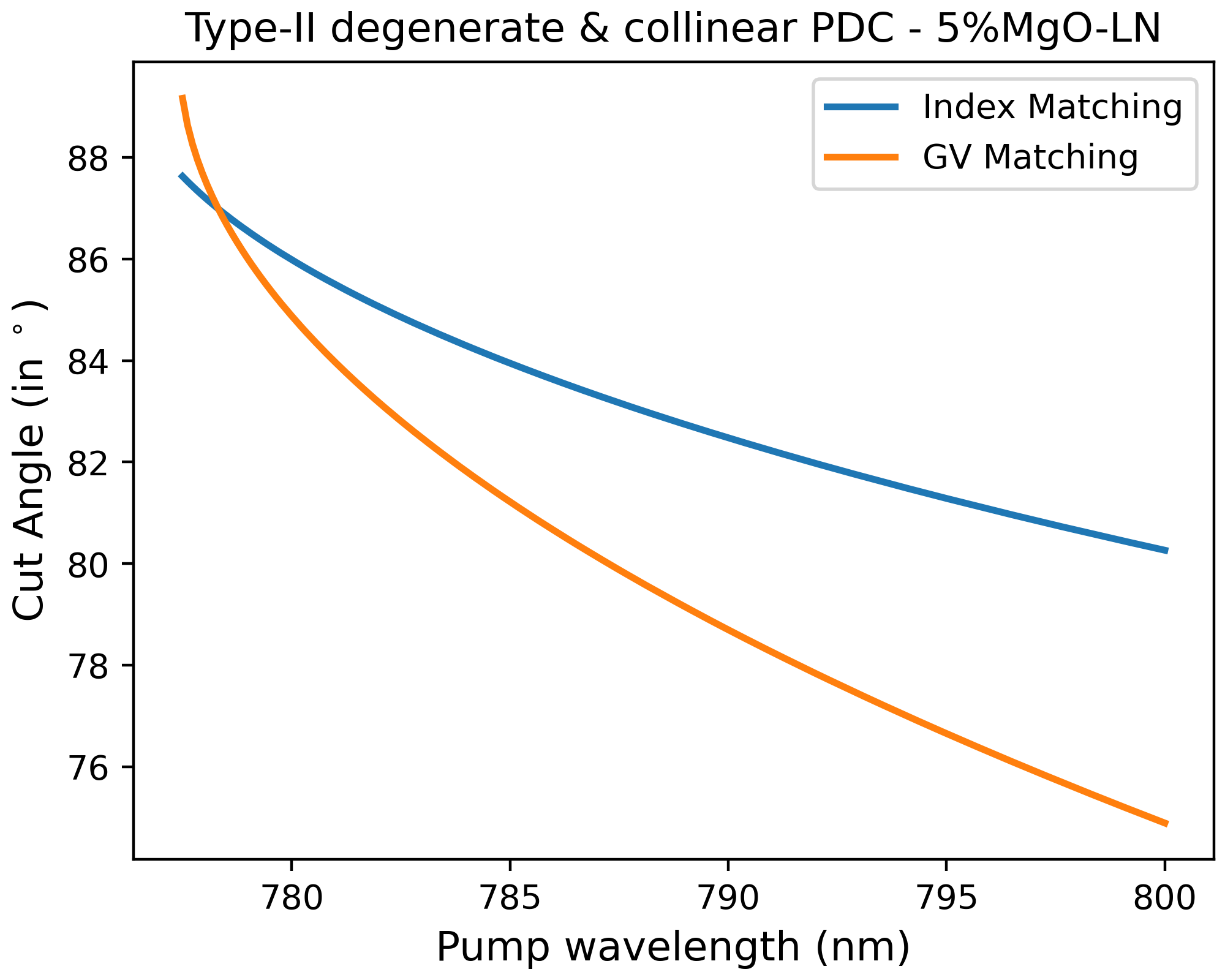}
\end{center}
\caption{Phase-matching and group-velocity-matching curves for degenerate collinear type-II PDC in $5\%$ MgO:LN at a crystal temperature of $15^\circ\mathrm{C}$, as a function of pump wavelength. This shows that both conditions are satisfied at pump wavelength of $778.32~\mathrm{nm}$ and a cut angle of $86.98^\circ$\label{fig:pm-gvm 5mgoln}}.

\vspace{-8mm}
\end{figure}

\begin{figure}[b]
 \vspace{3mm}
 \includegraphics[width=0.4\textwidth]{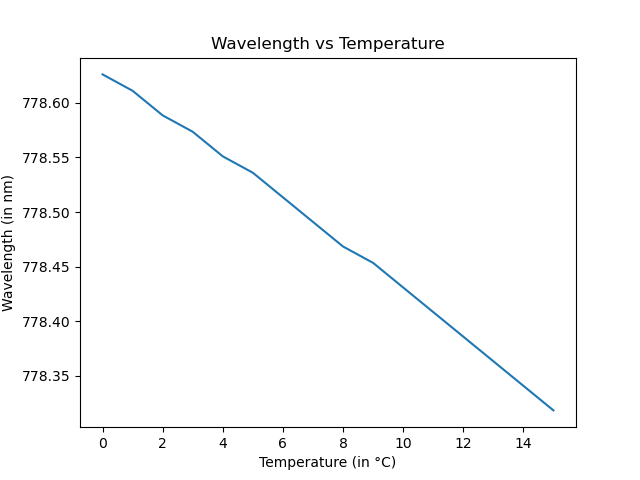}
\caption{The pump wavelength $\lambda_{GVM}$ as a function of crystal temperatures for the $5\%$ MgO:LN crystal.
\label{fig:wavelength vs temp 5mgoln}}

\vspace{-8mm}
\end{figure}

\begin{figure}
 \vspace{3mm}
 \includegraphics[width=0.4\textwidth]{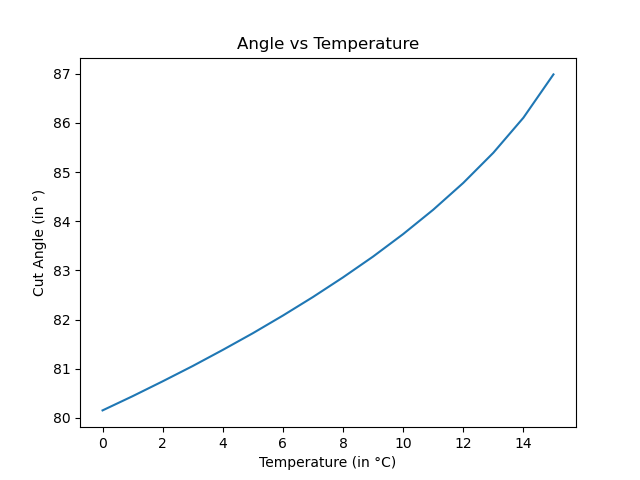}
\caption{Crystal cut angle, $\theta_{\mathrm{GVM}}$, as a function of crystal temperature for the $5\%$ MgO:LN crystal.
\label{fig:cutangle vs temperature 5mgoln}}
\vspace{-8mm}
\end{figure}
The simulation is continued for the $5 \%$ MgO doped LN over various temperature ranges. We found that the condition for the mode-selective addition is achieved at $15^ {\circ}\mathrm{C}$. At this temperature, the phase matching and group velocity matching curves intersect at $\lambda = 778~\mathrm{nm}$ wavelength for the pump photon. For the degenerate PDC process, the signal and idler photons are at the wavelength $\lambda_s =\lambda_i = 1556~\mathrm{nm}$. The phase matching and the group velocity matching angles are the same at this temperature and it is $\theta_{PM} = \theta_{GVM} =86.985^\circ$. The intersection of the phase matching and the group velocity matching curves in the $\lambda-\theta_{PM}$ plane is given in FIG~\ref{fig:pm-gvm 5mgoln}. The FIG.~\ref{fig:wavelength vs temp 5mgoln} and FIG.~\ref{fig:cutangle vs temperature 5mgoln} shows the variation of $\lambda_{GVM}$ and $\theta_{GVM}$ as a function of temperature. Although there is little variation in $\lambda_{\mathrm{GVM}}$, $\theta_{\mathrm{GVM}}$ changes by nearly $7^\circ$ as the temperature increases from $0^\circ\mathrm{C}$ to $15^\circ\mathrm{C}$. Beyond $15^\circ\mathrm{C}$, the phase-matching (PM) and GVM curves no longer intersect.

\begin{figure}[b]
\vspace{3mm}
\begin{center}
     \includegraphics[width=0.45\textwidth]{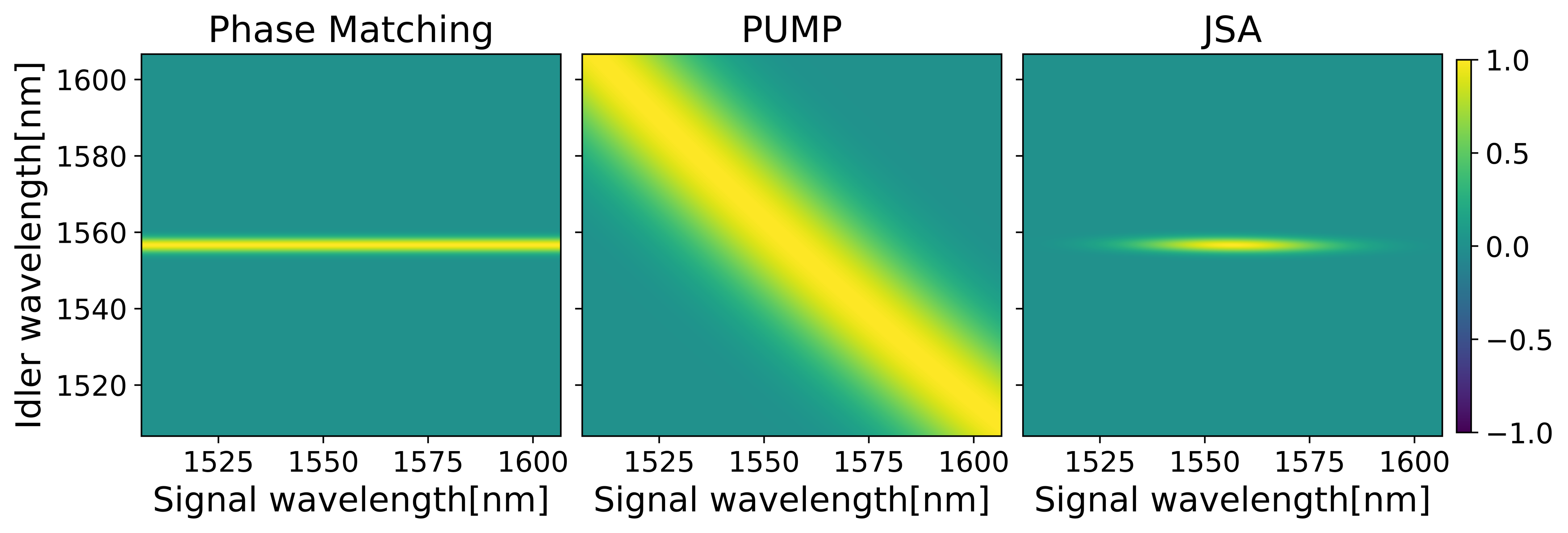}
\end{center}
\caption{Phasematching function (PMF) $\phi(\omega_s,\omega_i)$, pump envelope function (PEF) $\alpha(\omega_s,\omega_i)$, Joint Spectral Amplitude (JSA) for $5\%$ MgO:LN crystal at $15^\circ\mathrm{C}$. The pump wavelength is set to $\lambda_{GVM}=778.318~\mathrm{nm}$ with a zero order HG mode pump of FWHM bandwidth $10~\mathrm{nm}$ and the crystal length is $10~\mathrm{mm}$. The Schmidt number obtained from SVD is $K = 1.003$.
\label{fig:mode 0 jsa 5mgoln}}
\vspace{-8mm}
\end{figure}

\begin{figure}
\vspace{3mm}
 \includegraphics[width=0.4\textwidth]{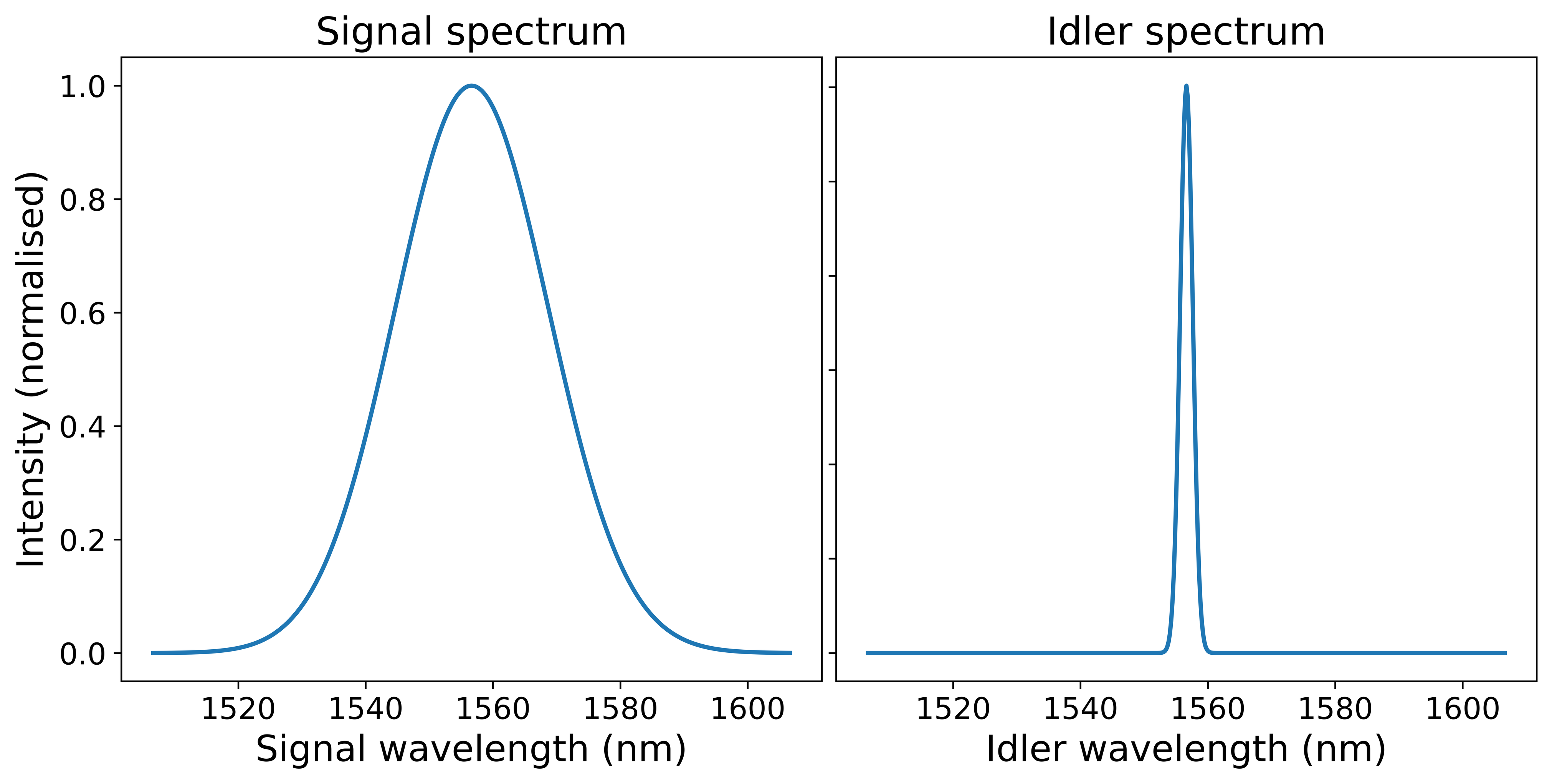}
\caption{The spectral modes of the signal and idler photons when the spectral mode of the pump is zeroth order HG mode.} 
\label{fig:mode 0 spectrum}
\vspace{-8mm}
\end{figure}

The simulated pump envelope function (PEF), phase matching function (PMF) and the joint spectral amplitude (JSA) are given in FIG~\ref{fig:mode 0 jsa 5mgoln}. Crystal length was set at $L=10~\mathrm{mm}$ and the pump FWHM bandwidth at $\sigma_{FWHM}=10~\mathrm{nm}$. The experimental parameters chosen according to the conditions discussed before resulted in a phase matching function that is horizontally oriented. This ensures that the mode selection happens in the signal photon while leaving the idler photon as a narrow gaussian function ~\cite{grice2001eliminating,namdar2025spectro}. The Schmidt number is calculated by performing the singular value decomposition on the joint spectral amplitude and it is $K=1.003$. The generated spectral modes of the signal and idler photons are given in FIG~\ref{fig:mode 0 spectrum}.

\begin{figure}
\vspace{3mm}
 \begin{center}
 \includegraphics[width=0.45\textwidth]{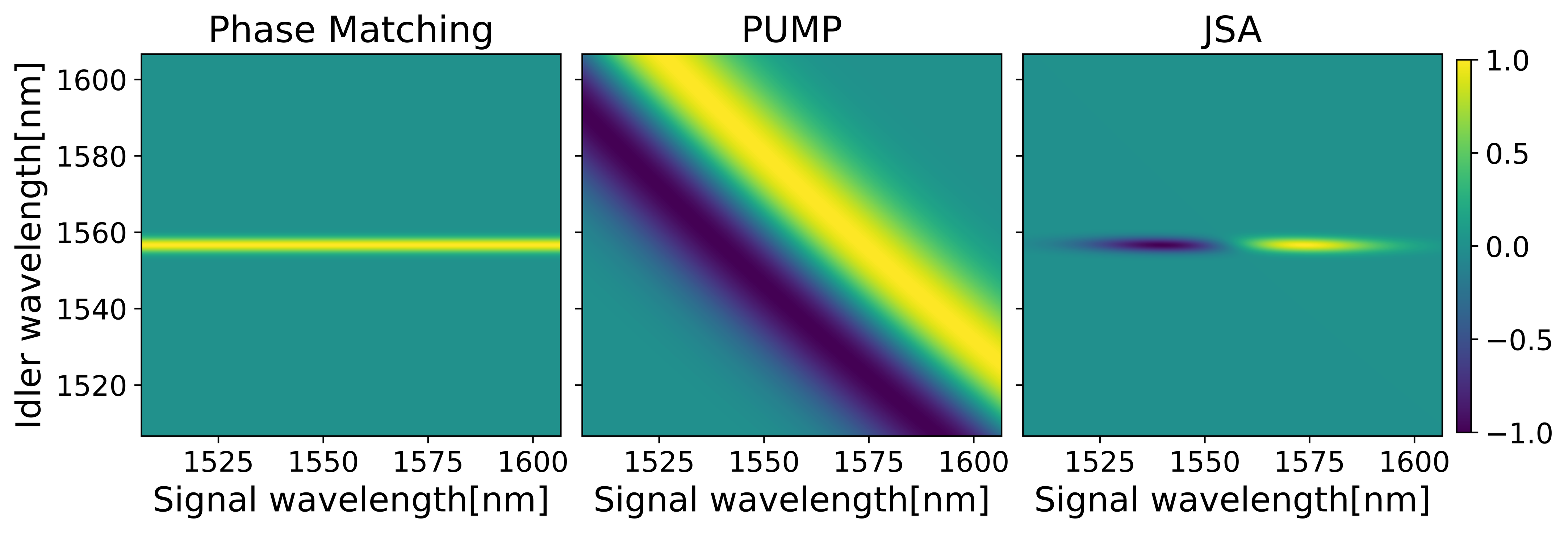}
  \end{center}
  \caption{The PMF, PEF and JSA for the $5\%$ MgO:LN crystal at $15^\circ\mathrm{C}$ with pump wavelength \(\lambda_{GVM}=778.318~\mathrm{nm}\). A first order HG mode with the FWHM bandwidth of \(10~\mathrm{nm}\) and the crystal length \(L=10~\mathrm{mm}\) are used in the calculation. The Schmidt number obtained from SVD is $K = 1.010$.}
\label{fig:mode 1 jsa 5mgoln}

\vspace{-8mm}
\end{figure}

\begin{figure}
\vspace{3mm}
 \includegraphics[width=0.4\textwidth]{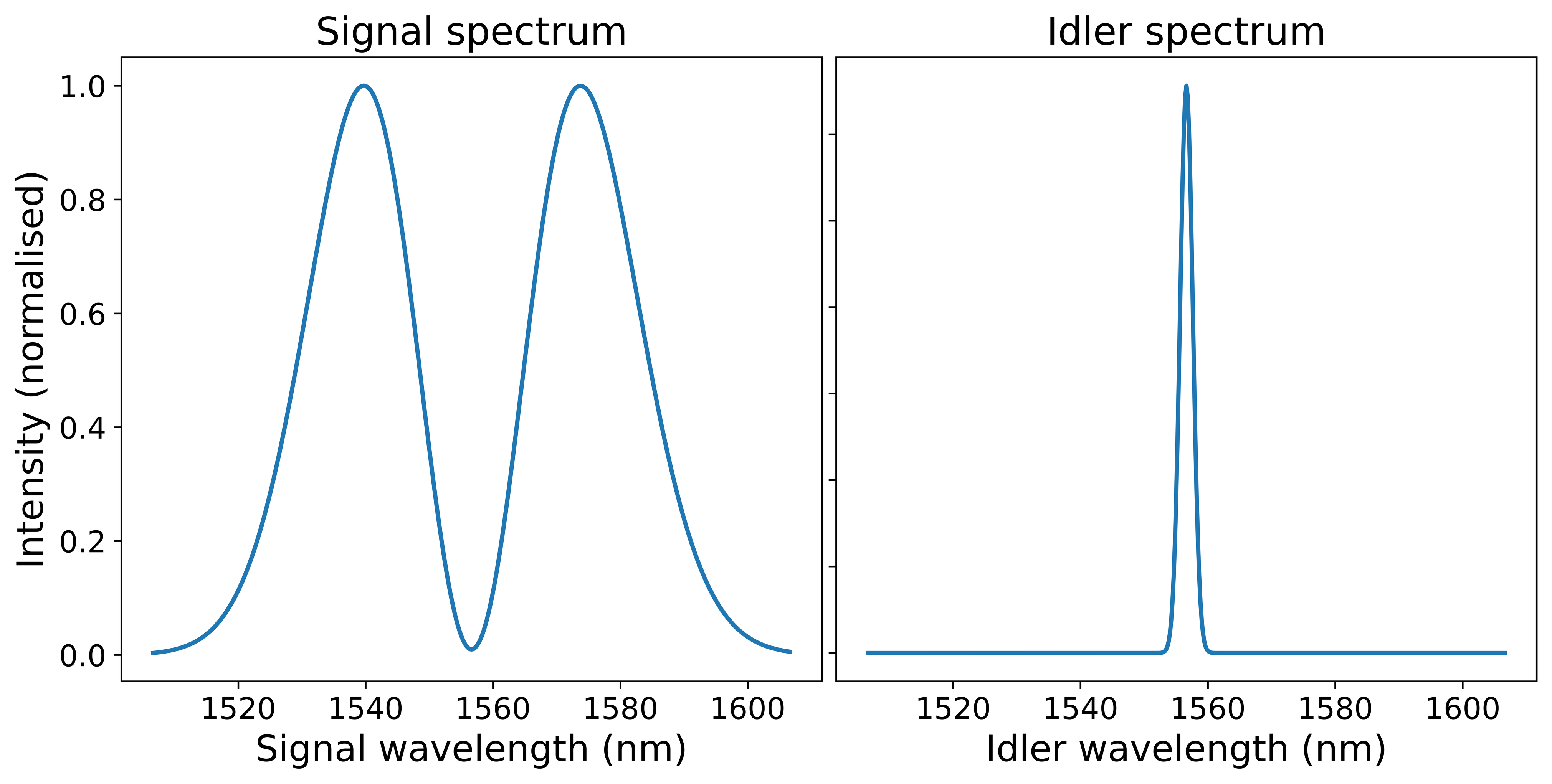}
\caption{The spectral modes of the signal and idler photons when the spectral mode of the pump is first order HG mode.}
\label{fig:mode 1 spectrum}
\vspace{-8mm}
\end{figure}

\begin{figure}
\vspace{3mm}
 \begin{center}
 \includegraphics[width=0.45\textwidth]{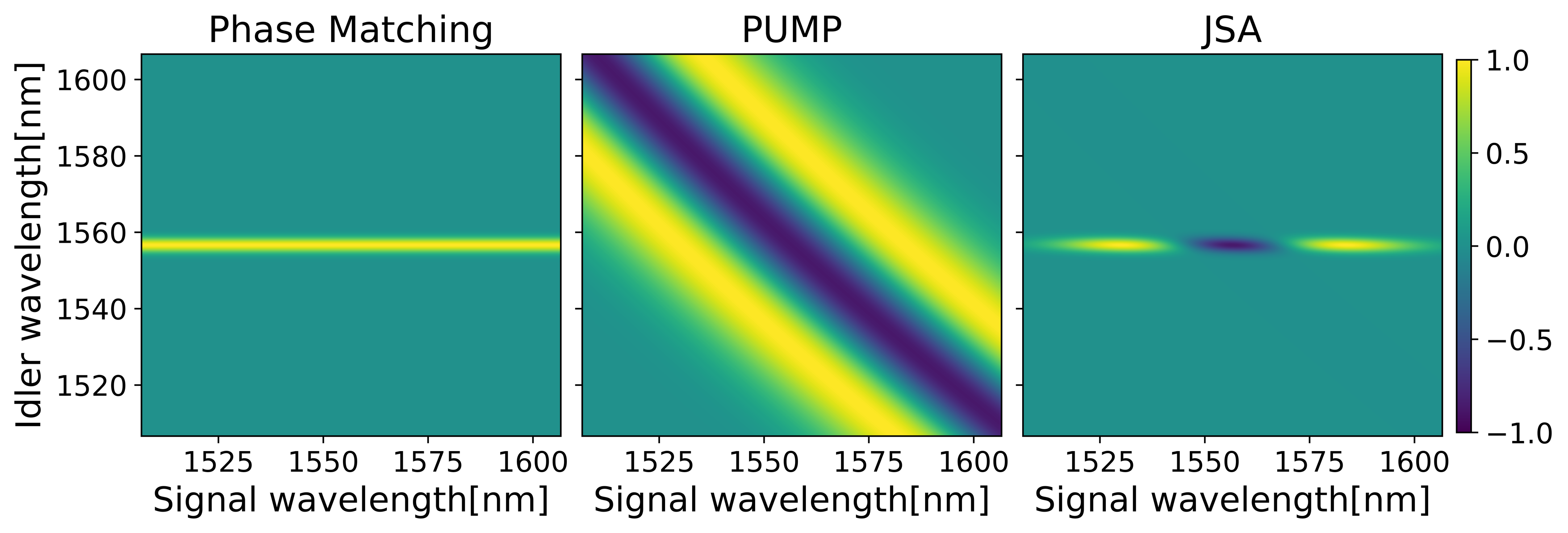}
\caption{The PMF, PEF and JSA for the $5\%$ MgO:LN crystal at $15^\circ\mathrm{C}$ with pump wavelength \(\lambda_{GVM}=778.318~\mathrm{nm}\). A second order HG mode with the FWHM bandwidth of \(10~\mathrm{nm}\) and the crystal length \(L=10~\mathrm{mm}\) are used in the calculation. The Schmidt number obtained from SVD is $K = 1.017$.}
\label{fig:mode 2 jsa 5mgoln}
 \end{center}
\vspace{-8mm}
\end{figure}
\begin{figure}
\vspace{3mm}
 \begin{center}
 \includegraphics[width=0.4\textwidth]{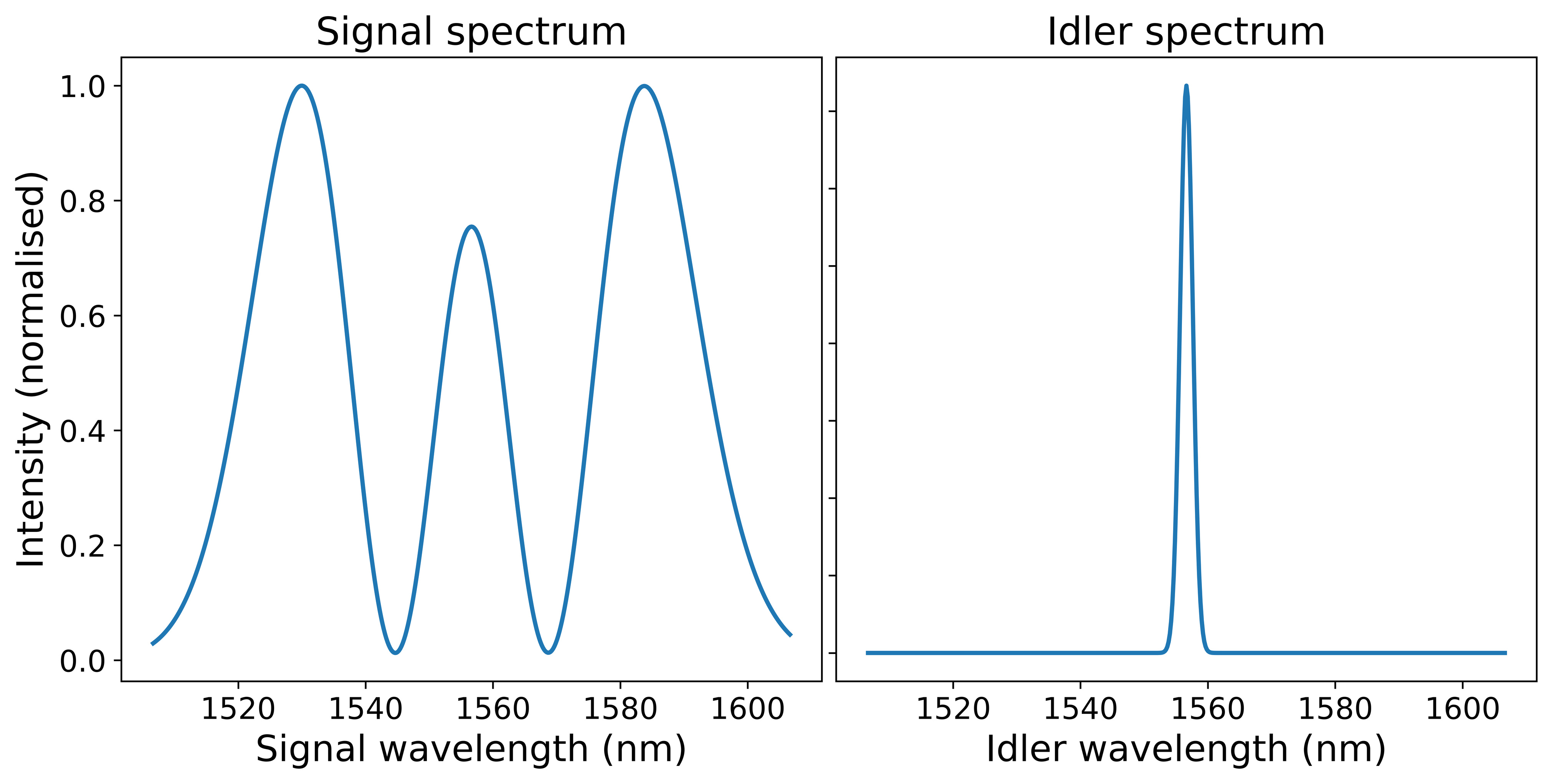}
\caption{The spectral modes of the signal and idler photons when the spectral mode of the pump is second order HG mode.
\label{fig:mode 2 spectrum}}
 \end{center}
\vspace{-8mm}
\end{figure}

 Since the condition for the mode-selective addition is achieved, the spectral mode of the signal is the same as that of the pump. The spectro-temporal mode of the pump can be shaped into any desired Hermite–Gauss ($HG$) mode, enabling the addition of the corresponding mode to the signal. The spectral mode of the idler is always the $HG_0$ mode. To add the $HG_1$ mode to the signal, we shape our pump to $HG_1$ mode. The PEF, PMF and JSA for this case are given in FIG~\ref{fig:mode 1 jsa 5mgoln} and Schmidt decomposition of the JSA revealed a K-value of $K = 1.010$. FIG~\ref{fig:mode 1 spectrum} shows the corresponding signal and idler modes. Similarly, for mode HG2, the calculated Schmidt number is $K=1.017$ and the plots for pump mode $HG_2$ are in FIG~\ref{fig:mode 2 jsa 5mgoln} and FIG~\ref{fig:mode 2 spectrum}.

As derived above, the GVM condition for achieving MSSPA does not impose a unique value on either the crystal length or the pump spectral bandwidth. Instead, the crystal length must be greater than a certain minimum value, or the pump spectral bandwidth must exceed a corresponding minimum value, as given by Eq.~\eqref{eq:long crystal broad pump}. Since the down-conversion process is collinear, the beam waist does not affect the mode-selective single-photon addition. However, the pump wavelength is subject to a stringent constraint, as the type-II collinear PDC process requires a specific pump wavelength to satisfy the phase-matching and GVM conditions for MSSPA. At $15^\circ C$ it was $\lambda_c = 778.318 ~nm$ with a Schmidt number of $K = 1.003$. At the same temperature of $15^\circ C$, the phase matching is possible for the pump wavelength range from $776.17~nm$ to $1330 ~nm$ for different phase matching angles. We studied the tolerance of the pump wavelength and identified that the Schmidt number K is $1.009$ at $776.17 ~nm$ and approaches $1.09$ at $786.3~nm$. Consequently, the tolerance in the pump wavelength  is approximately $10~nm$ to achieve MSSPA in $5\%$ MgO:LN. The simulation results presented here extend the possibility of performing the mode-selective SPA process with MgO:LN at the telecom wavelength.

\section{Single-photon addition using Type-I non-collinear parametric down-conversion process}
In the previous two sections, we presented the theoretical framework and simulation results of the mode-selective SPA process at telecommunication wavelengths using LN and its doped variants. We have demonstrated that mode-selective SPA can be achieved using MgO:LN at the telecommunication wavelength of \(1556\,\mathrm{nm}\). In the present section, we present the theoretical framework for the SPA using the noncollinear type-I PDC process given by AB Uren et al.~\cite{u2003photon}. This procedure establishes a set of conditions on the crystal length $L$, pump bandwidth $\sigma_p$ and the pump beam waist $w_0$ to generate spectrally pure downconverted photons thereby enabling the generation of the SPA without achieving the GVM between pump and signal photons. Consequently, SPA to multimode quantum field $\hat{\rho}$ can be realized at arbitrary wavelengths for which the refractive index matching is satisfied. This enables the generation of SPA at telecommunication wavelengths.  

The joint spectral amplitude (JSA) is a product of the pump envelope function and the phase-matching function. For the noncollinear PDC process, the phasematching function itself is a product of the transverse and the longitudinal phasematching functions. For a non-collinear and degenerate type-I PDC process $\omega_s=\omega_i=\frac{\omega_p}{2}$. The phase-matching condition for this process is given by

\begin{equation}\label{pm type1}
    n_e(\lambda_p,\theta_{PM})=n_o(\lambda_s)cos(\theta_s)
\end{equation}

Since it is a degenerate Type-I PDC process, due to the same polarization of the signal and idler photons, the emission angle of them will be the same $\theta_s=-\theta_i$. Consequently, the condition for SPA given in Eq.~\eqref{eq: uncorrelation condition} reduces to,

\begin{equation}\label{eq:uncorrelation type1}
\frac{1}{\sigma_p^{2}} + \frac{\gamma \left( K_{p}^{'} - K_{s}^{'}\cos\theta_{s} \right)^{2} L^{2}}{2}\  - \ \frac{ \left( K_{s}^{'}\sin\theta_{s} \right)^{2} w_{o}^{2}}{2} = 0
\end{equation}

For the purpose of approximation, let us consider a pump with bandwidth $\sigma_p$ broad enough to make the first term in the Eq.~\eqref{eq:uncorrelation type1} negligibly smaller than the next two terms. Rearranging the second and third terms of the equation establishes a relationship between the pump beam waist and the crystal length,

\begin{equation}\label{eq:w and L}
    \frac{w_{o}}{L} = \frac{\sqrt{\gamma}{(K}_{p}^{'} - K_{s}^{'}\cos\theta_{s})}{K_{s}^{'}\sin\theta_{s}}
\end{equation}

We assume that the first term in Eq.~\eqref{eq:uncorrelation type1} is smaller than the other two terms by using a broader pump bandwidth. However, there exists a lower limit on the pump bandwidth for this approximation to be valid. This lower limit can be obtained by substituting Eq.~\eqref{eq:w and L} into the joint spectral amplitude (JSA) given by Eq.~\eqref{eq:JSA},

\begin{equation}\label{eq:pump bandwidth}
    \sigma_{p} > \frac{1}{\sqrt{\gamma} L(K_{p}^{'} - K_{s}^{'}\cos\theta_{s})}
\end{equation}

Therefore, as long as the pump bandwidth exceeds the limit given in Eq.~\eqref{eq:pump bandwidth}, and the relationship between beam waist and crystal length is maintained along with the phase-matching conditions, it is possible to generate spectrally uncorrelated photons, which is essential for the SPA. 

There is one specific case, where the above condition fails and leads to a unrealistic condition. Let us assume the GVM is achieved between the pump and the signal photons are matched,
\begin{equation}\label{eq:gvm type1}
    K_{p}^{'} - K_{s}^{'}\cos\theta_{s} = 0
\end{equation}

Under this case equation (20) and (21) becomes,
\begin{align}\label{eq: type1 condition break point}
    \frac{w_{o}}{L} = 0 \quad \text{and} \quad
    \sigma_{p} > \infty
\end{align}
So when the GVM condition is satisfied, very unrealistic experimental conditions arise. From Eq.~\eqref{eq: type1 condition break point} it is clear that to achieve SPA the pump beam size should be zero or the crystal length should be infinite and the lower limit for the pump bandwidth should also be set to infinite. In the Type-II collinear PDC process, the group velocity matching between the pump and the signal photon is the necessary condition to achieve the single-mode addition. Whereas, as given in the Eq.~\eqref{eq:w and L} and Eq.~\eqref{eq:pump bandwidth}, SPA is possible for the Type-I noncollinear PDC process without achieving the group velocity matching condition. But there is a limitation for this procedure, as it provides unrealistic conditions when the group velocities are matched. 

Therefore, at this point in the phase matching range for which the GVM Eq.~\eqref{eq:gvm type1} is satisfied, the conditions for SPA Eq.~\eqref{eq:w and L} and Eq.~\eqref{eq:pump bandwidth} are irrelevant. So for this unique case, Eq.~\eqref{eq:uncorrelation type1} is revisited.
\begin{equation*}
\frac{1}{\sigma_p^{2}} + \frac{\gamma\left( K_{p}^{'} - K_{s}^{'}\cos\theta_{s} \right)^{2} L^{2}}{2}\  - \ \frac{\left( K_{s}^{'}\sin\theta_{s} \right)^{2}w_{o}^{2}}{2} = 0
\end{equation*}

Substituting Eq.~\eqref{eq:gvm type1} we have,
\begin{equation*}
\frac{1}{\sigma_p^{2}} - \ \frac{\left( K_{s}^{'}\sin\theta_{s} \right)^{2}w_{o}^{2}}{2} = 0
\end{equation*}
Rearranging this will result in,
\begin{equation}\label{eq:condition with gvm type1}
    \sigma_p = \frac{\sqrt{2}}{\left( K_{s}^{'}\sin\theta_{s} \right)w_{o}}
\end{equation}

Therefore, the present procedure for the SPA requires only the fulfilment of the phase-matching condition along with the conditions given in Eq.~\eqref{eq:w and L} and Eq.~\eqref{eq:pump bandwidth}. Under these conditions, the SPA is possible at arbitrary wavelength, wherever phase-matching is satisfied. An exception occurs when the GVM condition is also fulfilled. In this regime, the conditions given in Eq.~\eqref{eq:w and L} and Eq. ~\eqref{eq:pump bandwidth} lead to an unrealistic scenario. Instead, satisfying the condition given in Eq.~\eqref{eq:condition with gvm type1} alone is sufficient to realize the SPA process.

In the degenerate non-collinear type-I PDC process, the signal and idler photons are indistinguishable in every degree of freedom. This restricts us to achieve GVM matching between pump and signal alone, as the idler also identical to signal. Therefore, although SPA can still be realized in this configuration, mode-selective SPA cannot be achieved for this case. 

 \section{Simulations for the single photon addition using Type-I PDC process}
In the previous section, we presented the theoretical framework for realizing SPA with and without satisfying the GVM condition for a degenerate, noncollinear type-I PDC process. In this section, we present the simulation results for LN, $5\%$ MgO:LN and ZnO:LN  with the ZnO concentrations up to $8\%$, evaluated at various crystal temperatures using conditions given in Eq.~\eqref{eq:w and L} and ~\eqref{eq:pump bandwidth}. The analysis is primarily focuses on the realization of SPA at the telecommunication wavelength, particularly around $1550~nm$. In addition, we identify the pump wavelengths and the corresponding crystal cut angles for which the GVM condition and the phase matching condition are satisfied for type-1 non-collinear PDC simultaneously for the above mentioned crystals. The simulations for these wavelengths were performed using Eq.~\eqref{eq:condition with gvm type1}.


\begin{table*}[t]
\centering
\renewcommand{\arraystretch}{1.3}
\setlength{\tabcolsep}{14pt}
\begin{tabular}{c c c c c}

\hline\hline
& & \multicolumn{3}{c}{\textbf{Threshold pump spectral bandwidth}} \\
\textbf{Crystal} & \textbf{$\displaystyle \frac{w_0}{L} = \frac{
\sqrt{\gamma} \left(K_{p}^{'}-K_{s}^{'}\cos\theta_{s} \right)}{K_{s}^{'}\sin\theta_{s}}$} & $\sigma_{p}~(rad/s)$ & $\sigma_{p}$ ($nm$)& $\sigma_{FWHM}~(nm)=2\sqrt{2\ln2}~\sigma_p(nm)$ \\

\hline

LN & 0.2664 & $1.4243 \times 10^{13}$ & 4.5415 & 10.6943 \\
$5\%$ MgO:LN     & 0.2535 & $1.4993 \times 10^{13}$ & 4.7809 & 11.2581\\
$1\%$ ZnO:LN    & 0.1883 & $2.0081 \times 10^{13}$ & 6.4029 & 15.0775\\
$8\%$ ZnO:LN    & 0.1918 & $1.9664 \times 10^{13}$ & 6.2702 & 14.7651\\

\hline\hline
\end{tabular}

\caption{Comparison of the conditions required to realize the SPA using LN, $5\%$ MgO:LN and ZnO:LN crystals at temperature $T = 20^\circ\mathrm{C}$, as described by Eq.~\eqref{eq:w and L} and Eq.~\eqref{eq:pump bandwidth}. The numerical values listed in the comparison table are calculated for the $1~\mathrm{mm}$ long crystal with pump and signal wavelengths of $775~\mathrm{nm}$ and $1550~\mathrm{nm}$, respectively. The opening angle of the signal and idler photons with respect to the pump beam are $\theta_s=-\theta_i=2^\circ$.}
\label{tab:EPS_parameters1}
\end{table*}

\subsection{Simulations for SPA using noncollinear type-I PDC without achieving GVM}
The conditions to realize the SPA to a multimode quantum field without achieving the GVM between the pump and the signal photons are given in Eq.~\eqref{eq:w and L} and Eq.~\eqref{eq:pump bandwidth}. These conditions impose constraints on the crystal length $L$, pump beam waist $w_0$ and pump bandwidth $\sigma_p$. The realization of SPA requires the pump bandwidth $\sigma_p$ to exceed a minimum threshold value as given in Eq.~\eqref{eq:pump bandwidth}. Simulated numerical values to compute $\frac{w_0}{L}$ and the threshold value of the pump spectral bandwidth are given in Table~\ref{tab:EPS_parameters1} for LN, $5\%$ MgO:LN, $1\%$ ZnO:LN and $8\%$ ZnO:LN crystals at room temperature. The numerical values in the comparison table are calculated for the $1~\mathrm{mm}$ long crystal with pump and signal wavelengths of $775\,\mathrm{nm}$ and $1550\,\mathrm{nm}$, respectively. The opening angle of the signal and idler photons with respect to the pump beam is $\theta_s=-\theta_i=2^\circ$. 

\begin{figure}
\vspace{3mm}
 \begin{center}
 \includegraphics[width=0.45\textwidth]{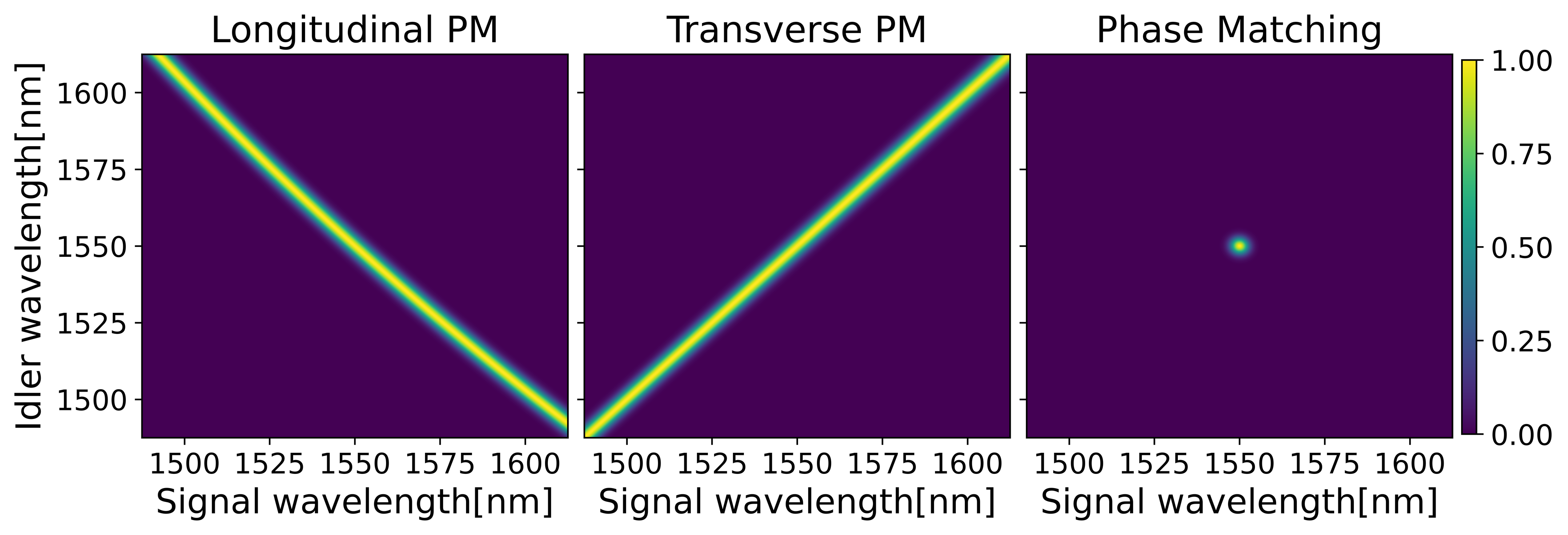}
\caption{Phase Matching Function of $5\%$ MgO:LN where pump wavelength is $775~\mathrm{nm}$, pump FWHM is $10~\mathrm{nm}$, crystal length is $10~\mathrm{mm}$, and crystal temperature is $20^\circ\mathrm{C}$.
\label{fig:775 mgo pm}}
 \end{center}
\vspace{-8mm}
\end{figure}

\begin{figure}
\vspace{3mm}
 \begin{center}
 
\includegraphics[width=0.45\textwidth]{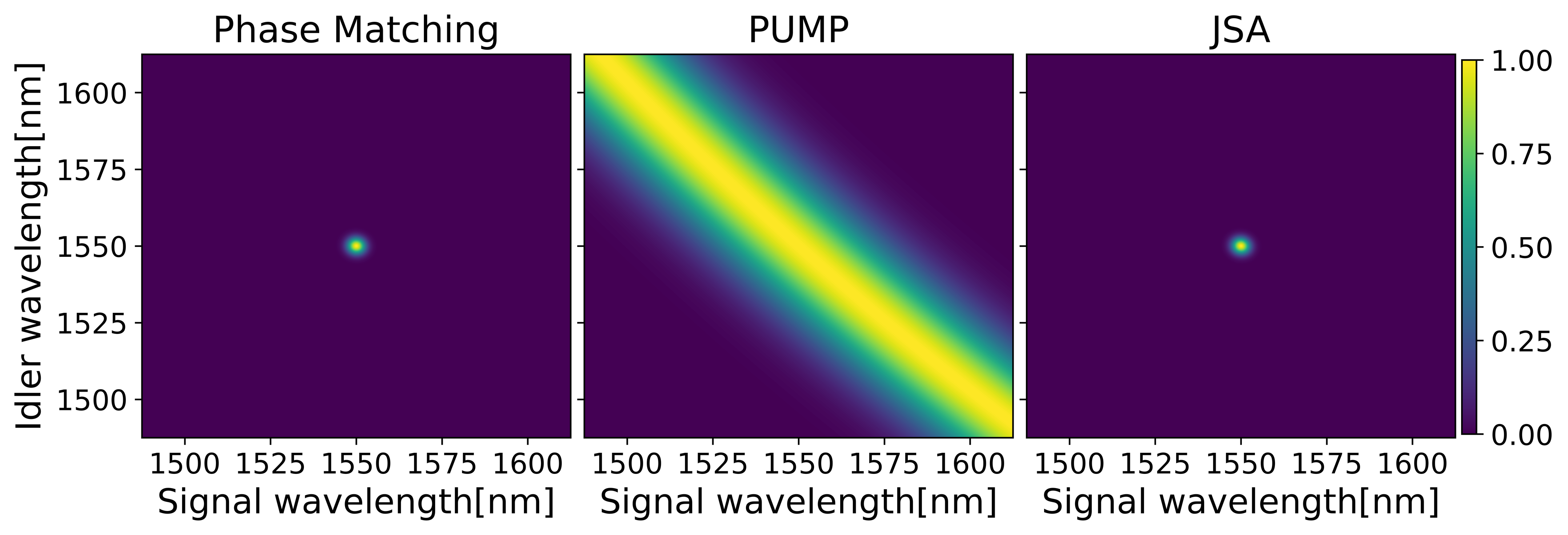}
\caption{Joint Spectral Amplitude of $5\%$ MgO:LN where pump wavelength is $775~\mathrm{nm}$, pump FWHM is $10~\mathrm{nm}$, crystal length is $10~\mathrm{mm}$, and crystal temperature is $20^\circ \mathrm{C}$. The Schmidt number calculated from SVD is $K = 1.00086$.
\label{fig:775 mgo jsa}}
 \end{center}
\vspace{-8mm}
\end{figure}

\begin{figure}
\vspace{3mm}
 \begin{center}
 
\includegraphics[width=0.45\textwidth]{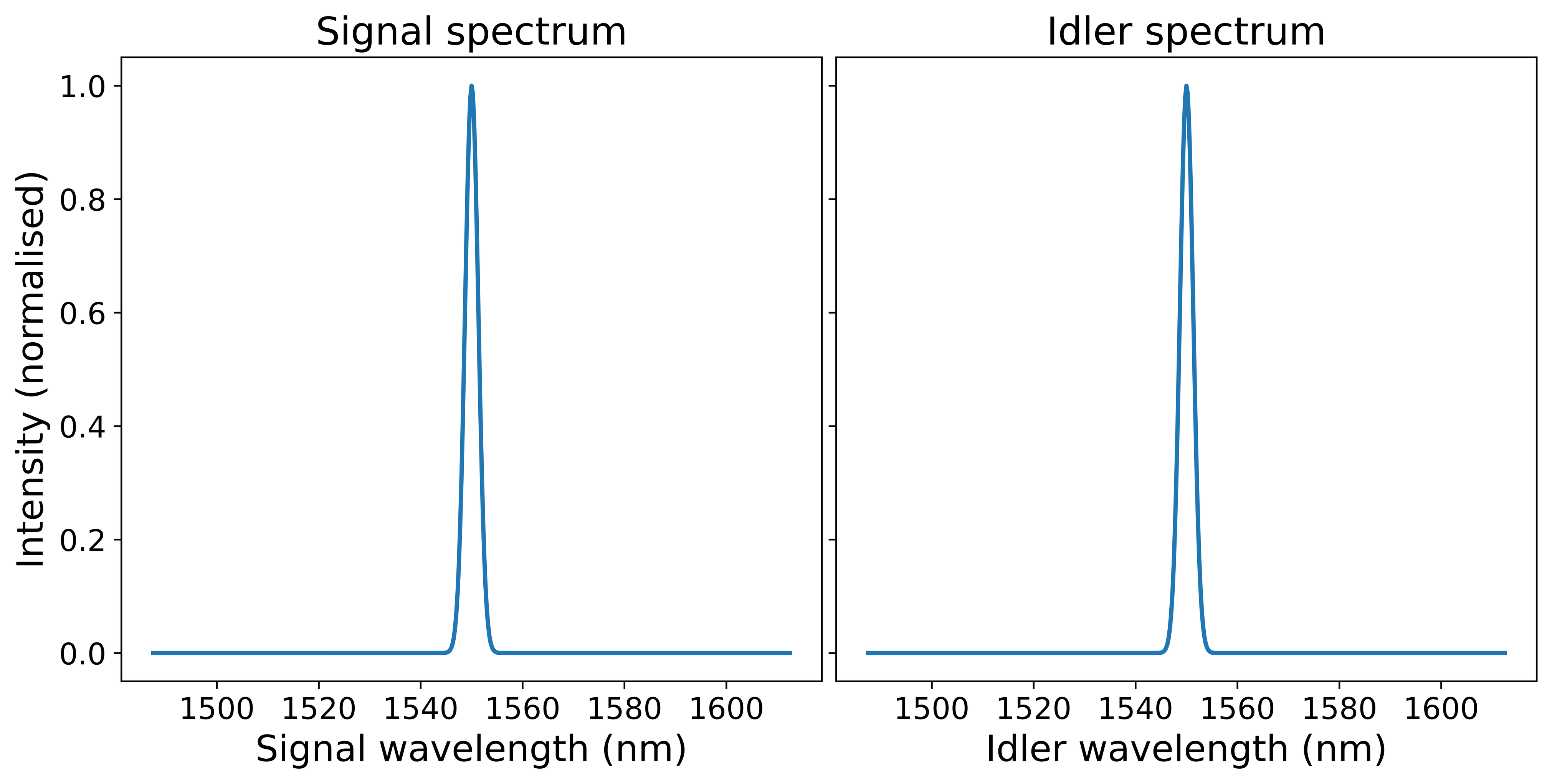}
\caption{Spectra of the signal and idler photons generated using $5\%$ MgO with a pump wavelength of $775~\mathrm{nm}$, a pump FWHM bandwidth of $10~\mathrm{nm}$, a crystal length of $10~\mathrm{mm}$, at $20^\circ~\mathrm{C}$.
\label{fig:775 mgo spectrum}}
 \end{center}
\vspace{-8mm}
\end{figure}

From Table~\ref{tab:EPS_parameters1}, let us choose the $5\%$ MgO:LN crystal for the simulation to realize SPA. As given in the table, the calculated values of $\frac{w_0}{L}$ and pump spectral bandwidth $\sigma_p$ are $0.2535$ and $4.7809 ~\mathrm{nm}$ respectively. These values are calculated for the $1~\mathrm{mm}$ long crystal at room temperature $T=20^\circ\mathrm{C}$ and half-cone angle $2^\circ$. Therefore, for a $1~ \mathrm{mm}$ long crystal, the pump beam waist is $w_0=253.5 ~\mathrm{\mu m}$ and the pump bandwidth must exceed $\sigma_p = 4.7809~ \mathrm{nm}$ (or $\sigma_{FWHM}=11.2581~\mathrm{nm}$) to realize the SPA. JSA simulations for pump wavelength of $775~\mathrm{nm}$, a down-conversion cone angle of $2^\circ$, and a crystal temperature of $20^\circ\mathrm{C}$ are performed for $5\%$ Mgo:LN. Under these conditions, the pump FWHM bandwidth, crystal length, and beam waist are chosen as $10~\mathrm{nm}$, $10~\mathrm{mm}$, and $2.535~\mathrm{mm}$, respectively. The simulated longitudinal PMF ($\phi_z(\omega_s,\omega_i)$), transverse PMF ($\phi_{\perp}(\omega_s,\omega_i)$), pump envelope function ($\alpha(\omega_s,\omega_i)$) and the JSA are shown in FIG.~\ref{fig:775 mgo pm} and FIG.~\ref{fig:775 mgo jsa}. Singular value decomposition is performed on the JSA.  The computed value of the Schmidt number is $K=1.00086$. The signal and idler eigenmodes are given in FIG.~\ref{fig:775 mgo spectrum}. These simulations are useful for identifying the optimal experimental conditions to realize SPA at telecommunication wavelengths without achieving the GVM between pump and signal photons.

In experimental implementations, it is difficult to maintain the exact parameter values specified above for the generation of single-photon-added states. Therefore, some tolerance in parameters such as the crystal length and beam diameter is necessary, while maintaining the Schmidt number $K$ within an acceptable range. Once the exact input parameters to achieve a Schmidt number as close as possible to 1 were calculated, they were altered a little bit to see how far these variables can be pushed away from their initial predictions and still maintain a reasonable Schmidt number $K =1.09$. For LN, we consider a pump wavelength of $775~\mathrm{nm}$, a down-conversion cone angle of $2^\circ$, and a crystal temperature of $20^\circ\mathrm{C}$. Under these conditions, the pump FWHM bandwidth, crystal length, and beam waist are chosen as $10~\mathrm{nm}$, $10~\mathrm{mm}$, and $2.66~\mathrm{mm}$, respectively. The SVD of the JSA yields a Schmidt number of $K = 1.00084$. To understand its experimental tolerance, only one variable was varied while the others were kept the same. We have identified that the $K<1.09$ was achieved between crystal lengths, $6.19~\mathrm{mm}$ and $14.7~\mathrm{mm}$. This implies that if the crystal length stays within this range, the Schmidt number will be under 1.09. Similarly, $K< 1.09$, when the beam waist size is between $1.759~\mathrm{mm}$ and $4.07~\mathrm{mm}$. In practical applications, this tolerance is more than sufficient. The condition on FWHM of pump is an inequality. Thus, it is subject only to a lower limit and is not restricted to a specific value, unlike the other parameters. Therefore, it does not pose a significant experimental constraint, provided that its value remains above the specified lower limit. Similarly, for 5\% MgO LN, the tolerance of crystal length and beam waist size was calculated around the values used for its JSA simulations earlier in this section. To maintain $K<1.09$, the crystal length has to be between $6.19~\mathrm{mm}$ and $14.7~\mathrm{mm}$ and the beam waist size has to be between $1.759~\mathrm{mm}$ and $4.07~\mathrm{mm}$. These tolerance ranges are very similar to that of LN.

\subsection{Simulation results of SPA under GVM conditions}

\begin{table*}[t]
\centering
\renewcommand{\arraystretch}{1.3}
\setlength{\tabcolsep}{14pt}
\begin{tabular}{c c c c c c }

\hline\hline
\textbf{Crystal} & $\lambda_{GVM} (nm)$ & $\theta_{GVM}$ & \multicolumn{3}{c}{\textbf{Pump spectral bandwidth}}\\
                 &                 &                & \(\sigma_p (rad/s)\) & \(\sigma_p (\mathrm{nm})\) & \(\sigma_{FWHM} (\mathrm{nm})\) \\
\hline

LN           & 1011.54 & 45.95 & $5.3664 \times 10^{13}$ & 17.1114 & 40.2739\\
$5\%$ MgO:LN & 1013.74 & 43.55 & $5.3748 \times 10^{13}$ & 17.1382 & 40.3570 \\
$1\%$ ZnO:LN & 929.22 & 49.41 & $5.3489 \times 10^{13}$ & 17.0557 & 40.1628\\
$8\%$ ZnO:LN & 931.98 & 48.65 & $5.3355 \times 10^{13}$ & 17.0131 & 40.0624\\

\hline\hline
\end{tabular}

\caption{Comparison table of the conditions required to realize the SPA when the GVM condition is satisfied. These numerical values are computed for LN, $5\%$ MgO:LN and ZnO:LN crystals at room temperature \(T=20^\circ~\mathrm{C}\) , as described by Eq.~\eqref{eq:condition with gvm type1}. The opening angle of the signal and idler photons with respect to the pump beam is $\theta_s=-\theta_i=2^\circ$ and the beam waist is $100 ~\mathrm{\mu m}$.}
\label{tab:EPS_parameters2}
\end{table*}

 The theoretical framework presented in the previous section provides a procedure for realizing SPA without satisfying the GVM condition. As discussed earlier, when the GVM condition between the pump and signal photons is satisfied, the conditions required by this procedure lead to an unrealistic experimental scenario. The condition given in Eq.~\eqref{eq:condition with gvm type1}, circumvents the problem under these circumstances. It identifies an optimal experimental condition for the beam waist $w_0$ and the spectral bandwidth $\sigma_p$ of the pump to realize SPA.

 Table~\ref{tab:EPS_parameters2} presents the comparison of the condition given in Eq.~\eqref{eq:condition with gvm type1} to optimize $w_0$ and $\sigma_p$ to realize SPA for the LN based crystals near the regime where GVM is satisfied. These numerical values are calculated for the pump and signal wavelengths for which the both the phase-matching and group velocity matching conditions were achieved. The beam waist of the pump is \(w_0 = 100 ~ \mathrm{\mu m}\) and the downconversion angle of signal and idler photons is $\theta_s=-\theta_i=2^\circ$.

Therefore, the general requirements for realizing SPA using the noncollinear type-I PDC process can be summarized as follows:
\begin{itemize}
    \item GVM is not satisfied: Use the conditions given in Eq.~\eqref{eq:w and L} and  Eq.~\eqref{eq:pump bandwidth} to realize SPA.
    \item GVM is satisfied: Use the condition given in Eq.~\eqref{eq:condition with gvm type1} to realize SPA. This condition is applicable over the wavelength range in which the GVM condition is approximately satisfied. 
\end{itemize} 

The conditions Eq.~\eqref{eq:condition with gvm type1} fill the gap left in the previous theoretical framework given by A.B. Uren et al, when the GVM Eq.~\eqref{eq:gvm type1} is satisfied.

\begin{figure}
\vspace{3mm}
 \begin{center}
 
\includegraphics[width=0.45\textwidth]{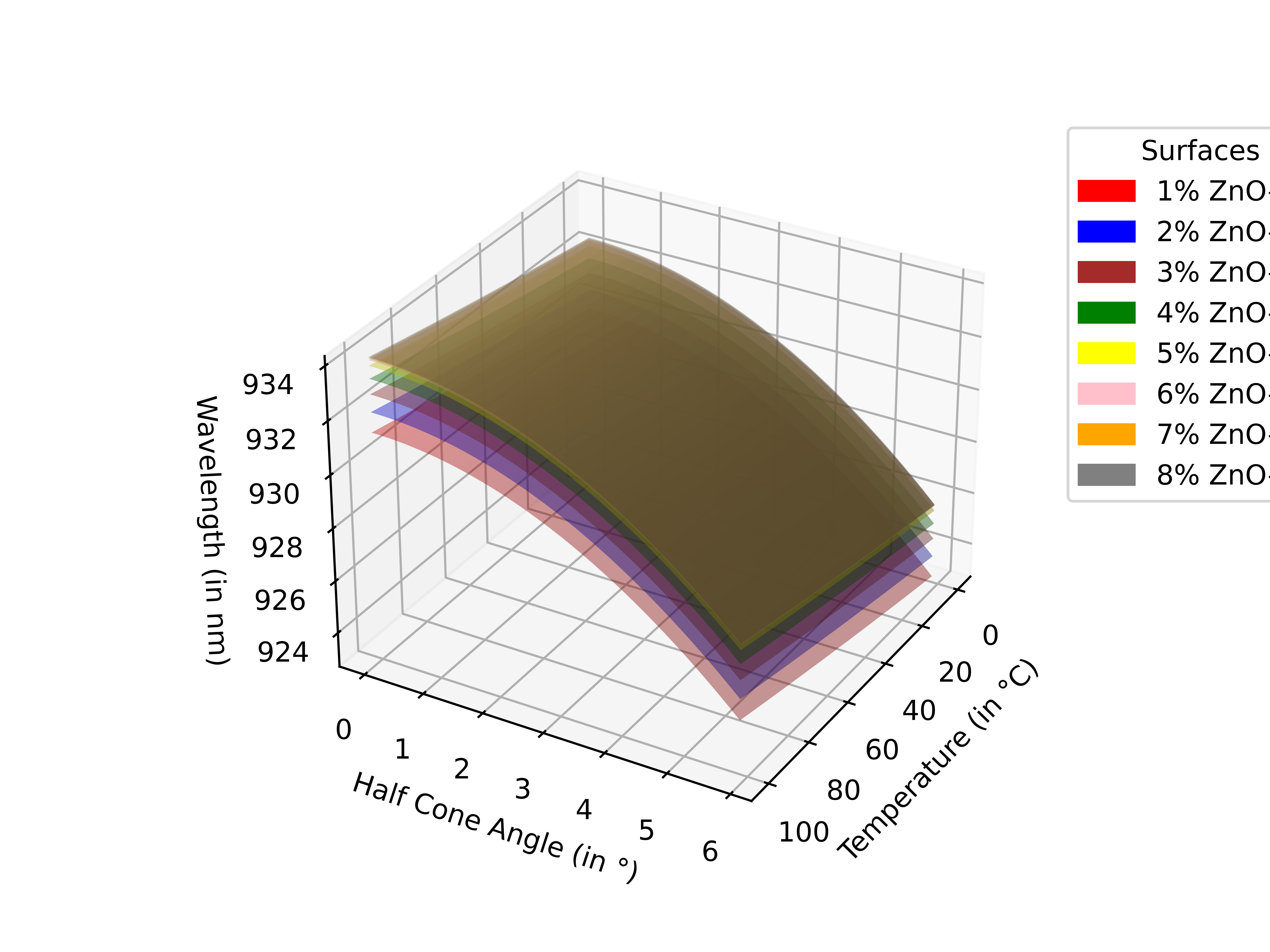}
\caption{Pump wavelength as a function of half-cone angle for different crystal temperatures and ZnO concentrations in ZnO:LN, under conditions satisfying both the phase-matching and GVM requirements.
\label{fig:wavelength 3d}}
 \end{center}
\vspace{-8mm}
\end{figure}

\begin{figure}
\vspace{3mm}
 \begin{center}
 
\includegraphics[width=0.45\textwidth]{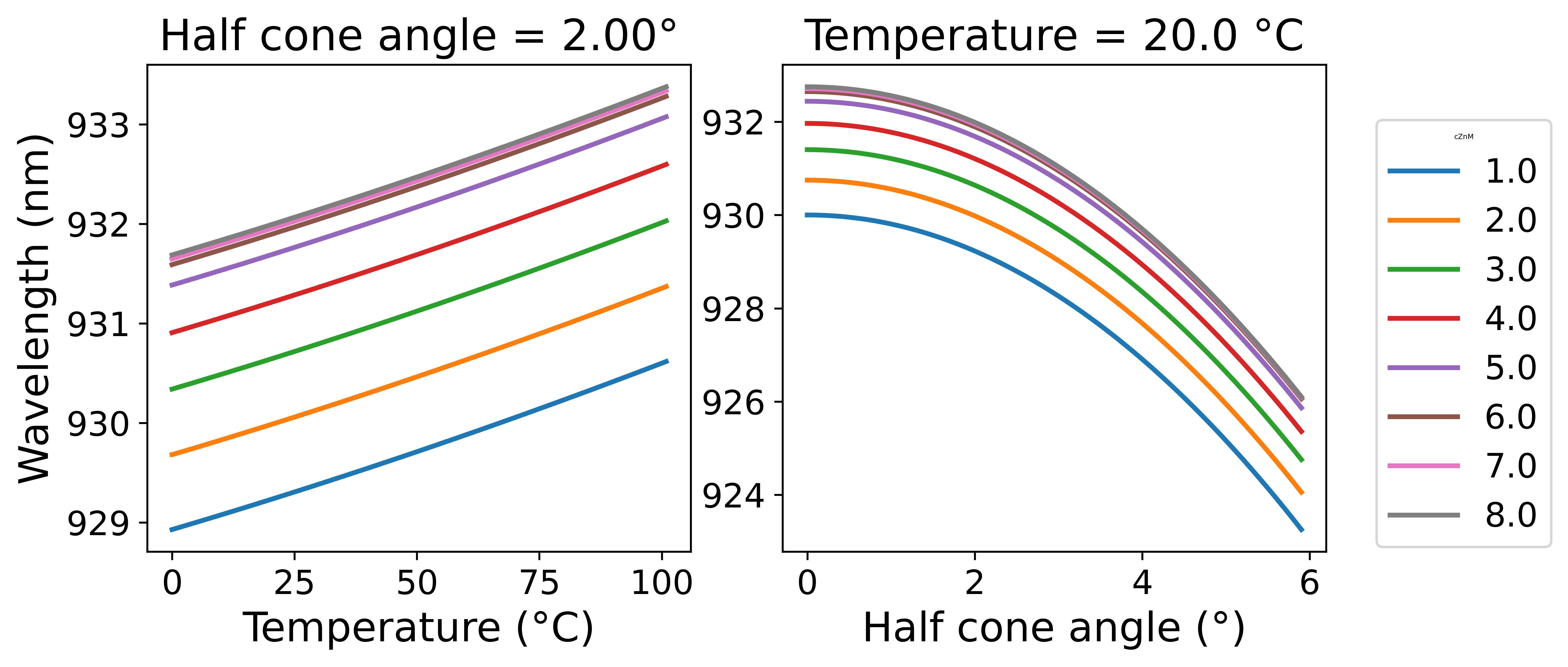}
\caption{Variation of the pump wavelength satisfying both the phase-matching and GVM conditions as a function of crystal temperature (left) and half cone angle (right) for different ZnO concentrations in LN.
\label{fig:wavelength combined}}
 \end{center}
\vspace{-8mm}
\end{figure}

\begin{figure}
\vspace{3mm}
 \begin{center}
 
\includegraphics[width=0.45\textwidth]{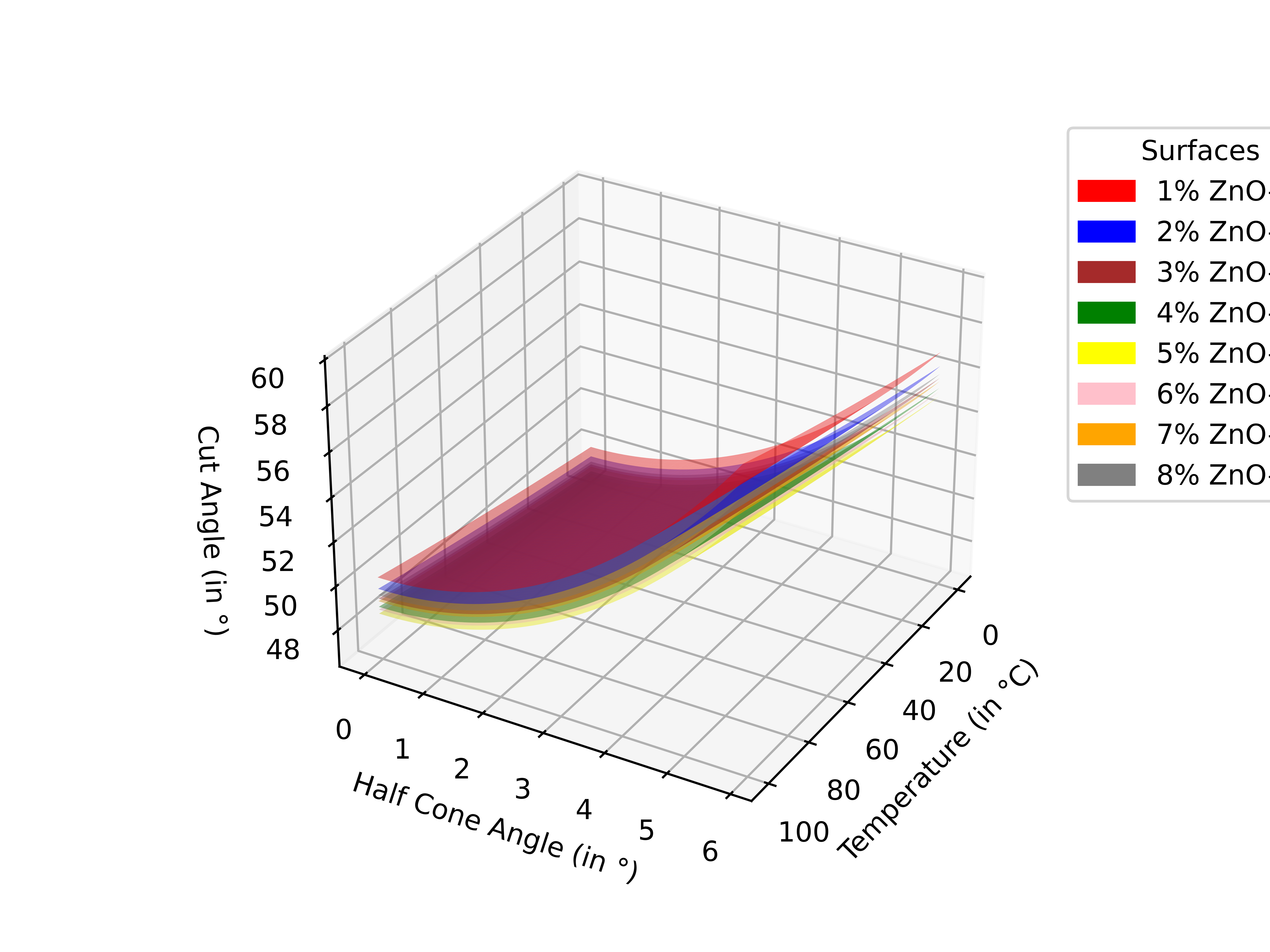}
\caption{Crystal cut angle plot for different half cone angles, temperatures and concentrations of ZnO:LN where the phase matching and GVM conditions are satisfied.
\label{fig:cutangle 3d}}
 \end{center}
\vspace{-8mm}
\end{figure}

\begin{figure}
\vspace{3mm}
 \begin{center}
 
\includegraphics[width=0.45\textwidth]{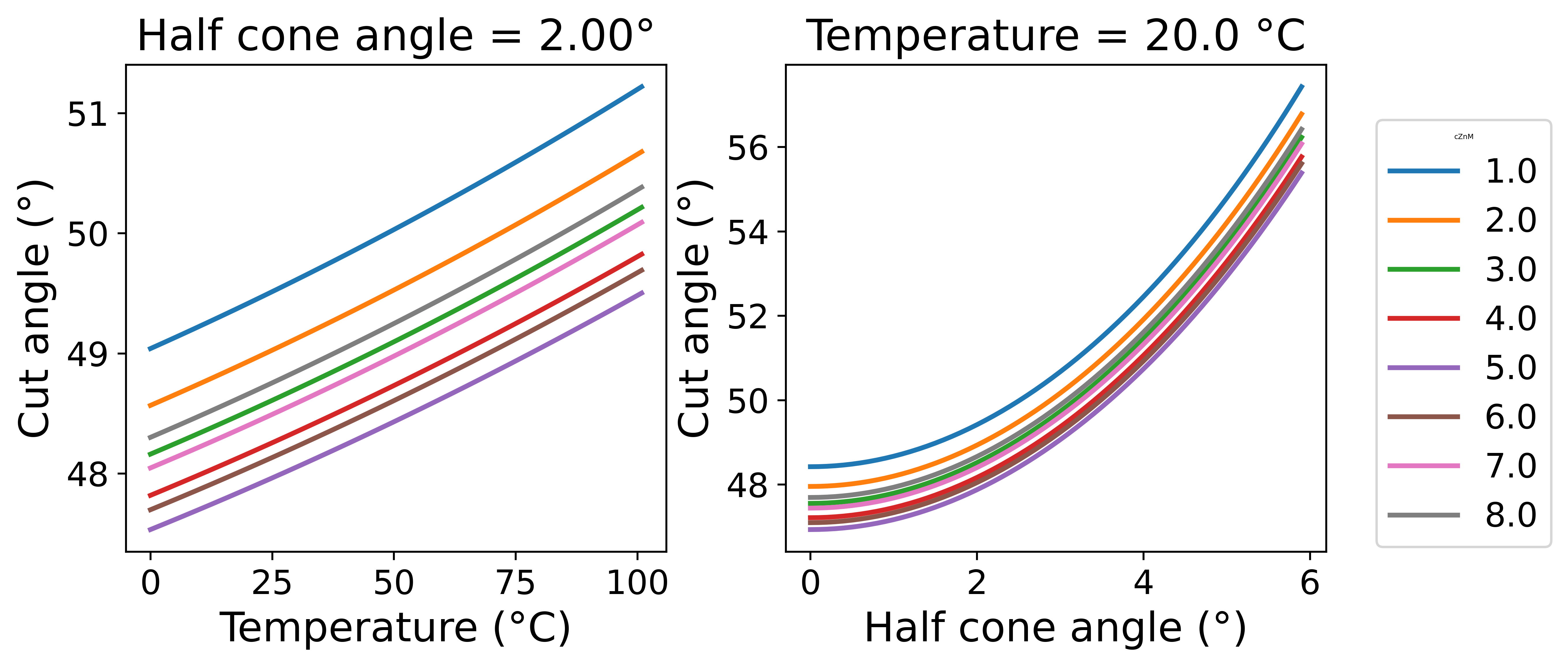}
\caption{Variation of crystal cut angles at which phase matching and GVM conditions are satisfied with varying  temperatures (left) and half cone angles (right) for different concentrations of ZnO in LN.
\label{fig:cut angle combined}}
 \end{center}
\vspace{-8mm}
\end{figure}

Further simulations are performed to identify the $\lambda_{GVM}$ and $\theta_{GVM}$ for various ZnO concentrations of the ZnO:LN crystal for a particular set of input parameters such as the crystal temperature $T$, down-conversion angle $\theta_s$ for which the PM and GVM are satisfied. The simulation results to identify $\lambda_{GVM}$ and $\theta_{GVM}$ values for a given $T$ and $\theta_s$ are given in FIG~\ref{fig:wavelength 3d} and FIG~\ref{fig:cutangle 3d}. The two-dimensional cross section of this simulation at $2^{\circ}$ downconversion angle and at crystal temperature $T = 20^{\circ}\mathrm{C}$ for $\lambda_{GVM}$ are given in FIG.~\ref{fig:wavelength combined}. The same for $\theta_{GVM}$ are given in FIG.~\ref{fig:cut angle combined}. Simulations were done for temperatures from $0^\circ\mathrm{C}$ to $100^\circ\mathrm{C}$ with a step size of $0.1^\circ\mathrm{C}$ and half cone angle from $0.1^\circ$ to $6^\circ$ in steps of $0.1^\circ$. 

\begin{figure}
 \vspace{3mm}
 \begin{center}
 \includegraphics[width=0.4\textwidth]{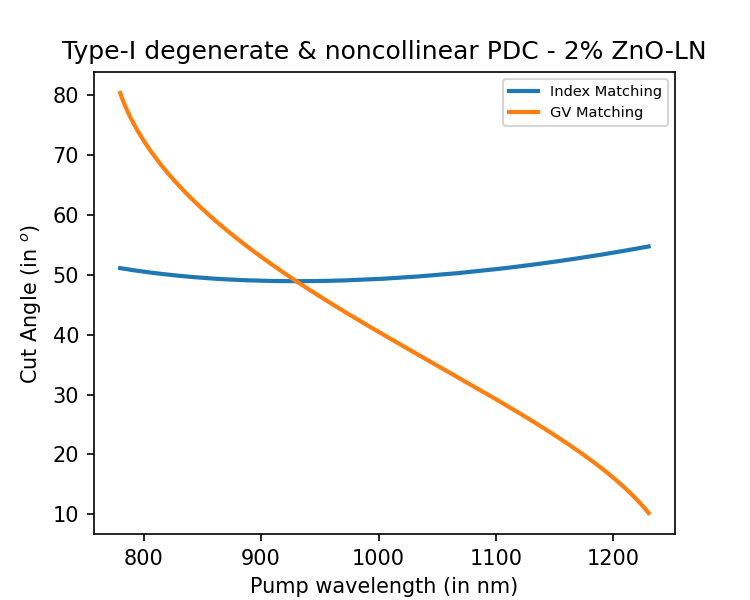}
\caption{Phase-matching and group-velocity-matching simulations at a crystal temperature of $20^\circ\mathrm{C}$ show that both conditions are simultaneously satisfied at a pump wavelength of $929.982~\mathrm{nm}$ and a cut angle of $48.93^\circ $.
\label{fig:932 zln pm-gvm}}
 \end{center}
\vspace{-8mm}
\end{figure}

\begin{figure}
\vspace{3mm}
 \begin{center}
 \includegraphics[width=0.45\textwidth]{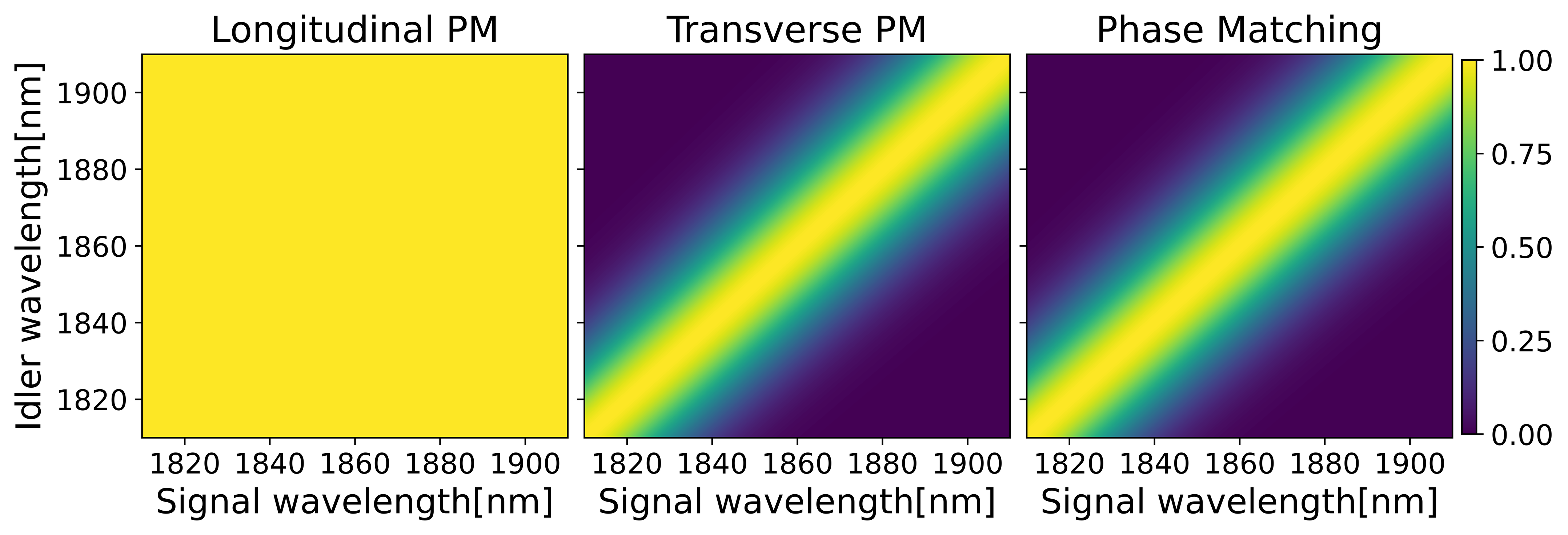}
\caption{Phase-matching function for $2\%$ ZnO:LN with a pump wavelength of $929.982~\mathrm{nm}$, pump FWHM of $10~\mathrm{nm}$, crystal length of $10~\mathrm{mm}$, and crystal temperature of $20^\circ\mathrm{C}$.
\label{fig:929 pm zln}}
 \end{center}
\vspace{-8mm}
\end{figure}

\begin{figure}
\vspace{3mm}
 \begin{center}
 
\includegraphics[width=0.45\textwidth]{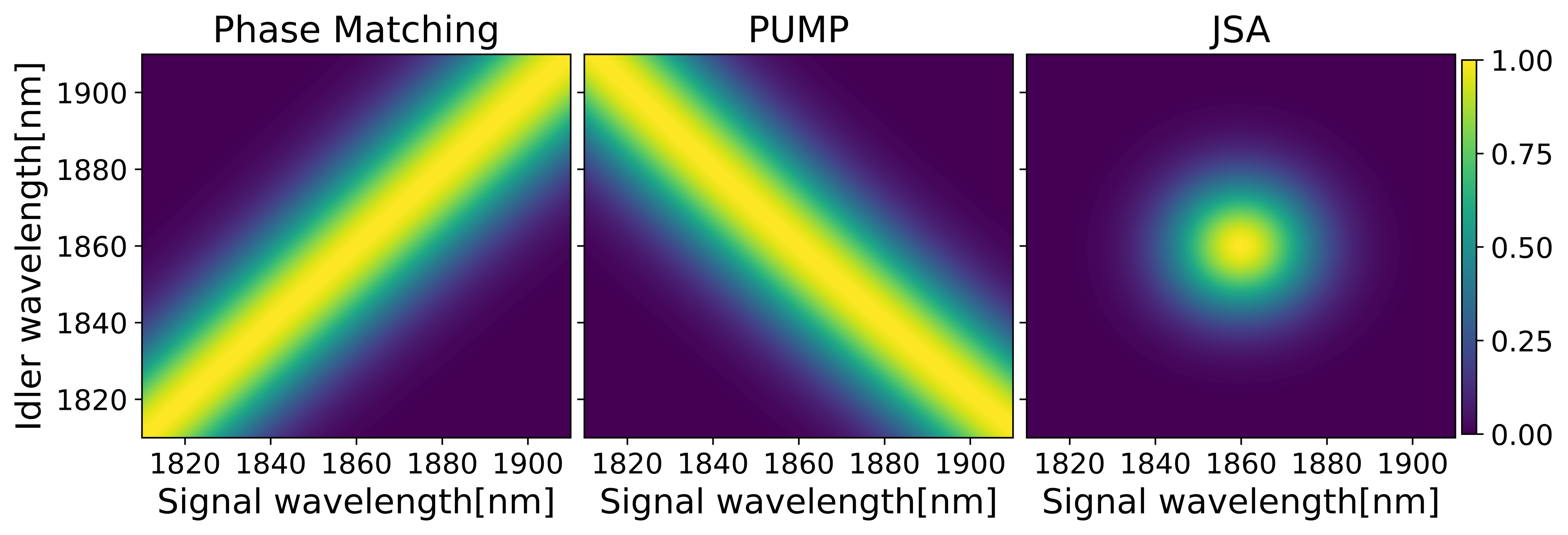}
\caption{Joint Spectral Amplitude of $2\%$ ZnO:LN with a pump wavelength of $929.982~\mathrm{nm}$, pump FWHM of $10 ~\mathrm{nm}$, crystal length is $10 ~\mathrm{mm}$, and crystal temperature is $20^\circ\mathrm{C}$. The Schmidt number obtained from the SVD is $K = 1.00002$. \label{fig:929 jsa zln}}
 \end{center}
\vspace{-8mm}
\end{figure}

\begin{figure}
\vspace{3mm}
 \begin{center}
 
\includegraphics[width=0.45\textwidth]{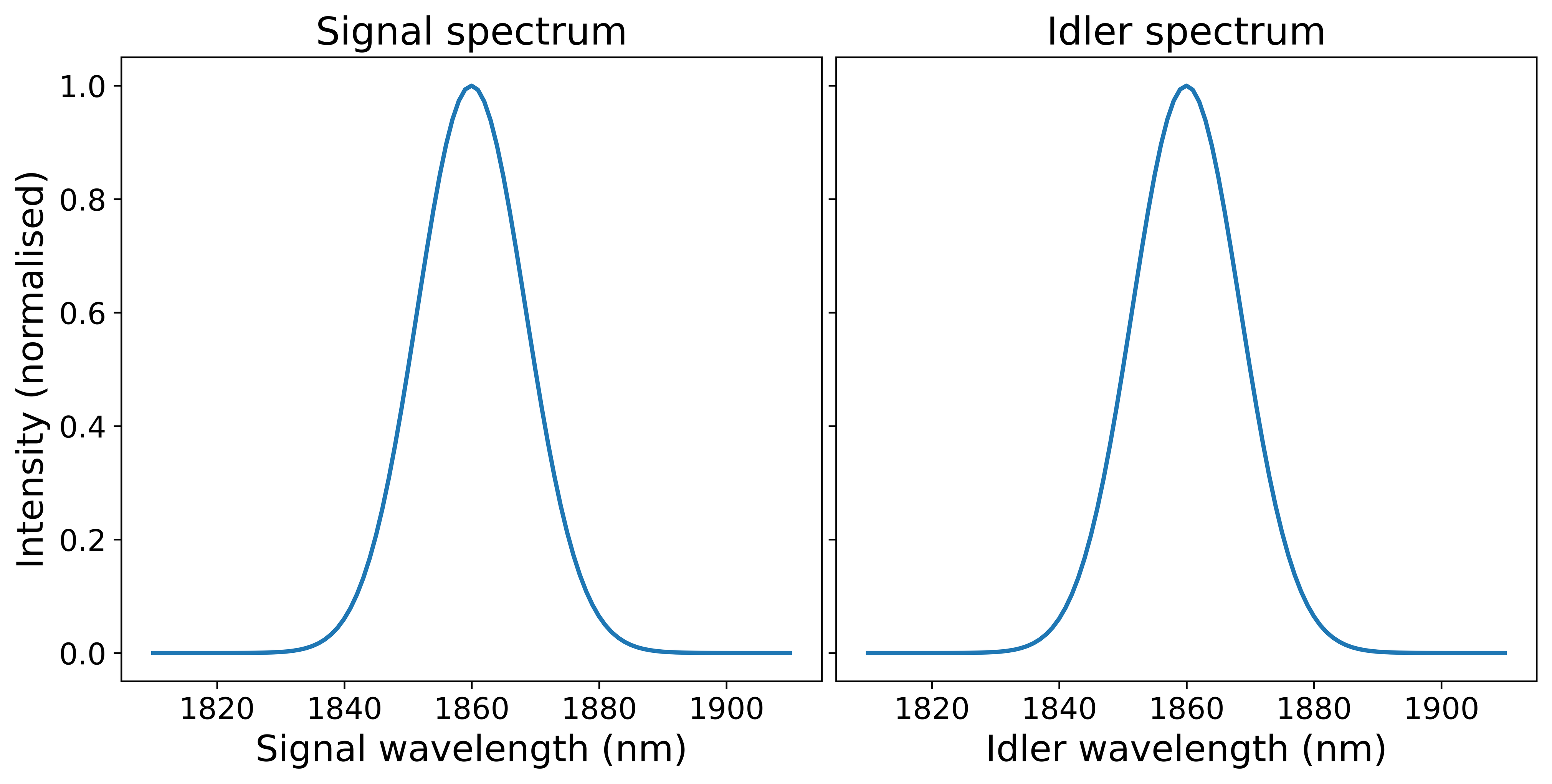}
\caption{Signal and idler spectra for $2\%$ ZnO:LN with a pump wavelength of $929.982~\mathrm{nm}$, pump FWHM of $10~\mathrm{nm}$, crystal length of $10~\mathrm{mm}$, and crystal temperature of $20^\circ\mathrm{C}$.
\label{fig:929 spectrum zln}}
 \end{center}
\vspace{-8mm}
\end{figure}

The simulation results for the PMF, PEF and JSA are given in  FIG.~\ref{fig:929 pm zln} and FIG.~\ref{fig:929 jsa zln} while satisfying the PM and GVM conditions simultaneously as given in FIG.~\ref{fig:932 zln pm-gvm}. A $10 ~\mathrm{mm}$ long ZnO:LN crystal at $T=20^{\circ}\mathrm{C}$ is used as the crystal parameters for the simulations. The pump wavelength, pump beam waist and the FWHM band width are $929.982~\mathrm{nm}$, $0.578~\mathrm{mm}$ and $10~\mathrm{nm},$ respectively. 
These parameters were chosen such that the conditions given in Eq. \eqref{eq:gvm type1} and Eq~\eqref{eq:condition with gvm type1} are satisfied. For these crystal and pump parameters, the calculated Schmidt number is $K=1.00002$. The corresponding signal and idler spectra are shown in FIG.~\ref{fig:929 spectrum zln}.

The GVM condition given by Eq.~\eqref{eq:gvm type1} makes the JSA independent of the crystal length. To investigate the robustness of the beam waist and pump bandwidth, we varied these parameters over the corresponding ranges. The Schmidt number remained below $K=1.09$ for beam waist values ranging from $0.380~\mathrm{mm}$ to $0.880~\mathrm{mm}$ and pump FWHM bandwidths ranging from $6.57~\mathrm{nm}$ to $15.21~\mathrm{nm}$.

\section{Discrepancies in the Reported Sellmeier Coefficients for LN}

\begin{figure}
\vspace{3mm}
 \begin{center}
 \includegraphics[width=0.45\textwidth]{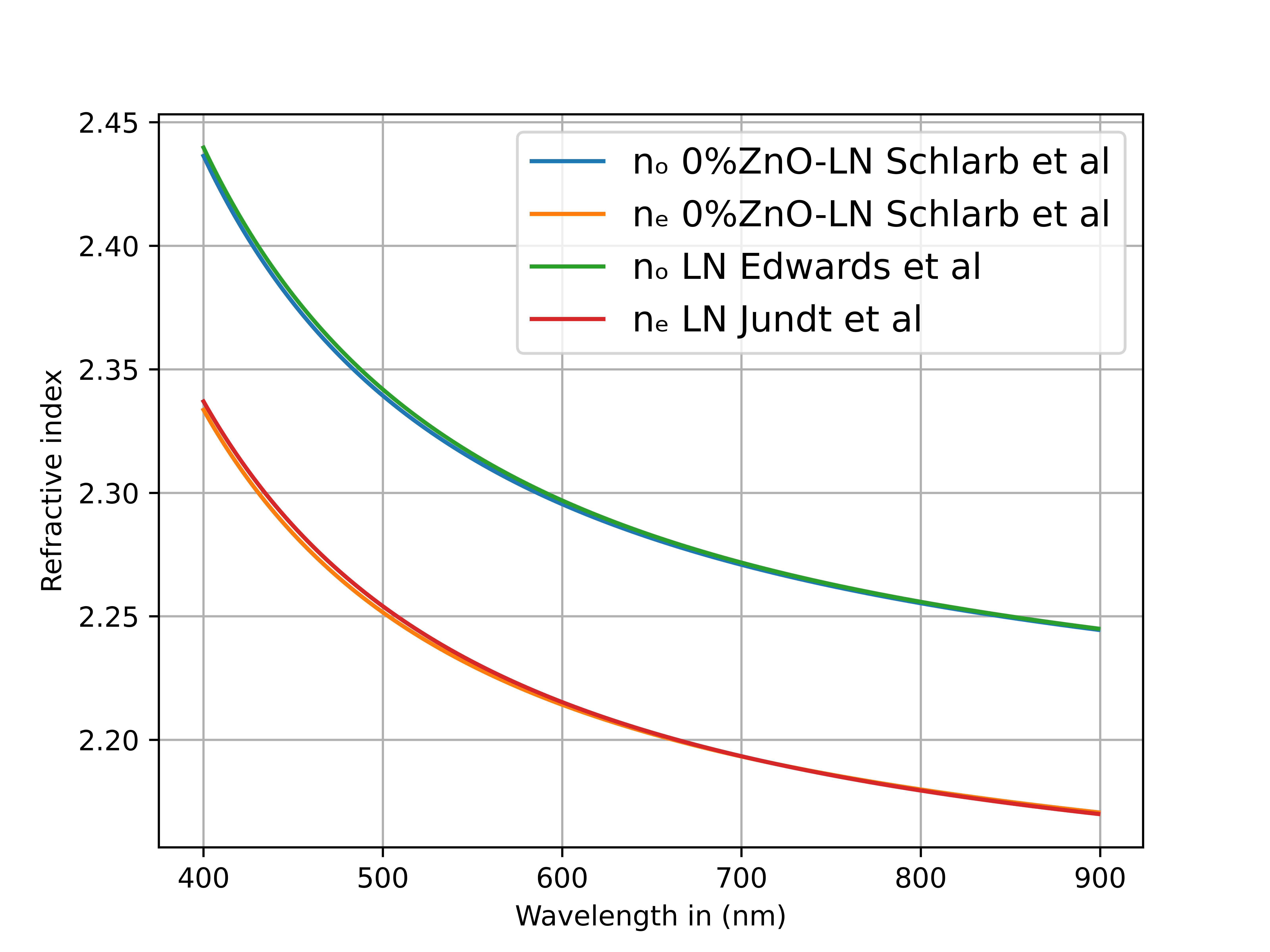}
\caption{ Refractive index curves for LN and ZnO:LN at a crystal temperature of $20^\circ$C, calculated using the coefficients reported by Edwards et al. for $n_o$ of LN ~\cite{edwards1984temperature}, Jundt et al. for $n_e$ of LN ~\cite{jundt1997temperature}, and Schlarb et al. for ZnO:LN ~\cite{schlarb1995refractive}.
\label{fig:LN dispersion curve}}
 \end{center}
\vspace{-8mm}
\end{figure}
The numerical simulation for the LN, $5\%$ MgO:LN and various concentrations of ZnO:LN crystal are presented. From the table - II, the group velocity matching wavelength $\lambda_{GVM}$ for the LN crystal is \(\lambda_{GVM}=1011.54~ \mathrm{nm} \). This value is calculated using the Sellmeier coefficient presented by Jundt et al~\cite{jundt1997temperature} and Edwards et al~\cite{edwards1984temperature}. The group velocity matching wavelength calculated for the $0\%$ ZnO:LN crystal would be expected to match with this value. However, the group velocity matching wavelength calculated using the Sellmeier equation given in Schlarb et al~\cite{schlarb1995refractive}, for the $0\%$ concentration is \(\lambda_{GVM}=929.09~ \mathrm{nm} \). The plot for the ordinary and extraordinary refractive indices for various wavelengths are given in Fig~\ref{fig:LN dispersion curve}. A very small mismatch in the refractive indices are observed between the two Sellmeier models. This leads to a major drift in the $\lambda_{GVM}$ value for LN. 
\section{Conclusion}
In this work, we presented a theoretical and numerical investigation focused on identifying the optimal parameters for realizing single-photon addition (SPA) to spectro-temporal multimode quantum fields at telecommunication wavelengths, with particular emphasis on $1550$ nm. We investigated two approaches: mode-selective single-photon addition (MSSPA) using collinear type-II PDC and SPA using noncollinear type-I PDC.

 We presented a detailed theory for the generation of MSSPA using the collinear type-II PDC process. We identified optimal phase-matching parameters for MSSPA and found that MSSPA is possible only for the $5\%$ MgO:LN crystal for the wavelength $1556$ nm at $15^\circ C$. We further extended our calculations for various crystal temperatures ranging from $0^\circ C$ to $15^\circ C$. By optimizing the crystal length and the pump bandwidth, we tailored the JSA to achieve the Schmidt number $K=1.003$.
 
 We have also identified optimal phase-matching parameters and experimental conditions to realize SPA using a noncollinear type-I PDC process. We presented an analytical expression to overcome the limitation posed by the A. U'Ren et al. work. By satisfying this analytical condition, we have presented the result for SPA with maximum purity given by the Schmidt number $K=1.00002$. Through our theoretical analysis, we identified optimal parameters such as crystal length, pump bandwidth, beam waist and phase matching angle for LN, $5\%$ MgO:LN and ZnO:LN crystals to realize SPA at $1550$ nm. We believe that this work will be helpful for the generation of non-Gaussian multimode spectro-temporal modes at telecommunication wavelengths.  

Overall, this work establishes a theoretical framework and identifies experimentally relevant conditions for generating spectro-temporal multimode single-photon-added quantum states at telecommunication wavelengths. These results may facilitate the generation and investigation of non-Gaussian multimode quantum states of light for applications in continuous-variable quantum information processing and photonic quantum technologies.

\begin{acknowledgments}
The authors acknowledge the support provided by DIAT (DU) through the Grant-in-Aid Program. The authors also thank Kanaka Raju Pandiri and Joseph Amalnathan (School of Quantum Technology, DIAT, Pune, India) for useful discussions.
\end{acknowledgments}


\bibliography{apssamp}

\end{document}